\documentclass{ws-procs9x6}
\usepackage{subfigure}

\newcommand{\kps}{\ensuremath{\,\mathrm{km/s}}}
\newcommand{\keV}{\ensuremath{\,\mathrm{keV}}}
\newcommand{\Gcc}{\ensuremath{\,\mathrm{GeV/\mathit{c}^2}}}
\newcommand{\cmm}{\ensuremath{\,\mathrm{cm^2}}}

\begin{document}

\title{An Introduction to Dark Matter Direct Detection Searches \& Techniques}

\author{T. Saab}

\address{Department of PhysicsUniversity Department, University of Florida,\\
Gainesville, FL 32611-8440, USA\\
$^*$E-mail: tsaab@uufl.edu}

\begin{abstract}
Weakly Interacting Massive Particles (WIMPs), are a leading candidate for the dark matter that is observed to constitute $\sim$25\% of the total mass-energy density of the Universe. The direct detection of relic WIMPs (those produced during the early moments of the Universe's expansion) is at the forefront of active research areas in particle astrophysics with a numerous international experimental collaborations pursuing this goal. This paper presents an overview of the theoretical and practical considerations common to the design and operation of direct detection experiments, as well as their unique features and capabilities.
\end{abstract}


\bodymatter


\section{Introduction}\label{sec:Intro}

The experimental evidence for the existence of dark matter as well as the theoretical description of the candidate particles are well described elsewhere in these proceedings. Suffice it to say that the $\Lambda-$CDM model of the Universe is very good at explaining the observed Universe on both the large (horizon) and small (galaxies) scales and sizes in between. Despite some difficulties with numerical simulations properly reproducing the dark matter substructure within galactic halos, the generally accepted model of galaxy structures is that of a baryonic component (e.g. a central bulge and spiral disk) embedded in a fairly smooth spherical halo of dark matte particles. The properties of the dark matter halo are inferred from the rotational kinematics of the baryons 
with the average velocity of the earth through the halo being $v_0 = 220\kps$, the velocity dispersion of the WIMPs at the Earth's location $v_{rms} = 270\kps$, the WIMP escape velocity $v_{esc}= 650\kps$, and the local WIMP density $\rho = 0.3\Gcc$.
With that information in hand one can ask the question: ``How often will a dark matter particle at Earth's location in the halo interact with a particular nucleus, and what is the expected distribution of recoil energies imparted to the nucleus?''.  The ability to formulate a sensible answer based on knowledge of the astrophysics, particle, and nuclear properties of the system in question is the basis of all direct detection experiments which aim to detect halo dark matter and study their properties based on the observation of WIMP interactions in their detectors.

In the following sections I will outline the broad theoretical considerations involved in calculating the expected rate and spectrum of interactions in a detector (\sref{sec:InteractionRate}), how properties of different detector material can be exploited to search for such interactions and what tricks and techniques are available to suppress background rates or disentangle the dark matter component in the presence of a given background (\sref{sec:BackgroundRejection}). I will then describe a representative set of experimental implementations (\sref{sec:Experiments}) before summarizing that status of the dark matter direct detection field (\sref{sec:Limits}).

\section{WIMP-Nucleus Interactions}\label{sec:InteractionRate}

Given the low velocity and subsequent momentum transfer in iterations between the halo WIMPs and nuclei in an earth based experiment, the dominant interaction mechanism is expected to be elastic scattering due to spin-independent interactions. The differential recoil energy spectrum of such an interaction between a WIMP of mass $m_\chi$ and a nucleus of mass $m_N$ (per unit detector mass) is given by \eref{eq:DiffRate1}~\cite{Jungman:283685,Lewin:1996vn}:
\begin{equation}
\frac{dR}{dQ}  =  \frac{\sigma_0 \rho_0}{\sqrt{\pi}v_0 m_\chi m_r^2} \, F^2(Q) \, T(Q)
\label{eq:DiffRate1}
\end{equation}
where $\rho_0$ is the WIMP density in the local neighborhood within the galactic halo, $\sigma_0$ is the elastic scattering cross-section between the WIMP and nucleus, $m_r$ is the WIMP-nucleus reduced mass: $m_r = \left(\frac{m_\chi m_N}{m_\chi + m_N}\right)$, $F(Q)$ is the nuclear form factor, and $T(Q)$ is a dimensionless integral over the local WIMP velocity distribution.
For spin-independent scattering, $\sigma_0$ is approximately proportional to the WIMP-proton scattering cross-section $\sigma_{\chi-p}$:  $\sigma_0\propto A^2 \sigma_{\chi-p}$. This relationship emphasizes the dependence of the interaction rate on the size of the nucleus $A$.
Of course, one could consider other mechanisms that may contribute to the interaction cross-section such spin-dependent interaction terms. In that case, the dependence on the target's nuclear properties is proportional to the total spin $J$ rather than $A^2$. All of the other factors relevant to calculating the interaction rate are unchanged.

\begin{figure}[!ht]
\begin{center}
\subfigure[][]{
\includegraphics[width=0.45\linewidth, keepaspectratio]{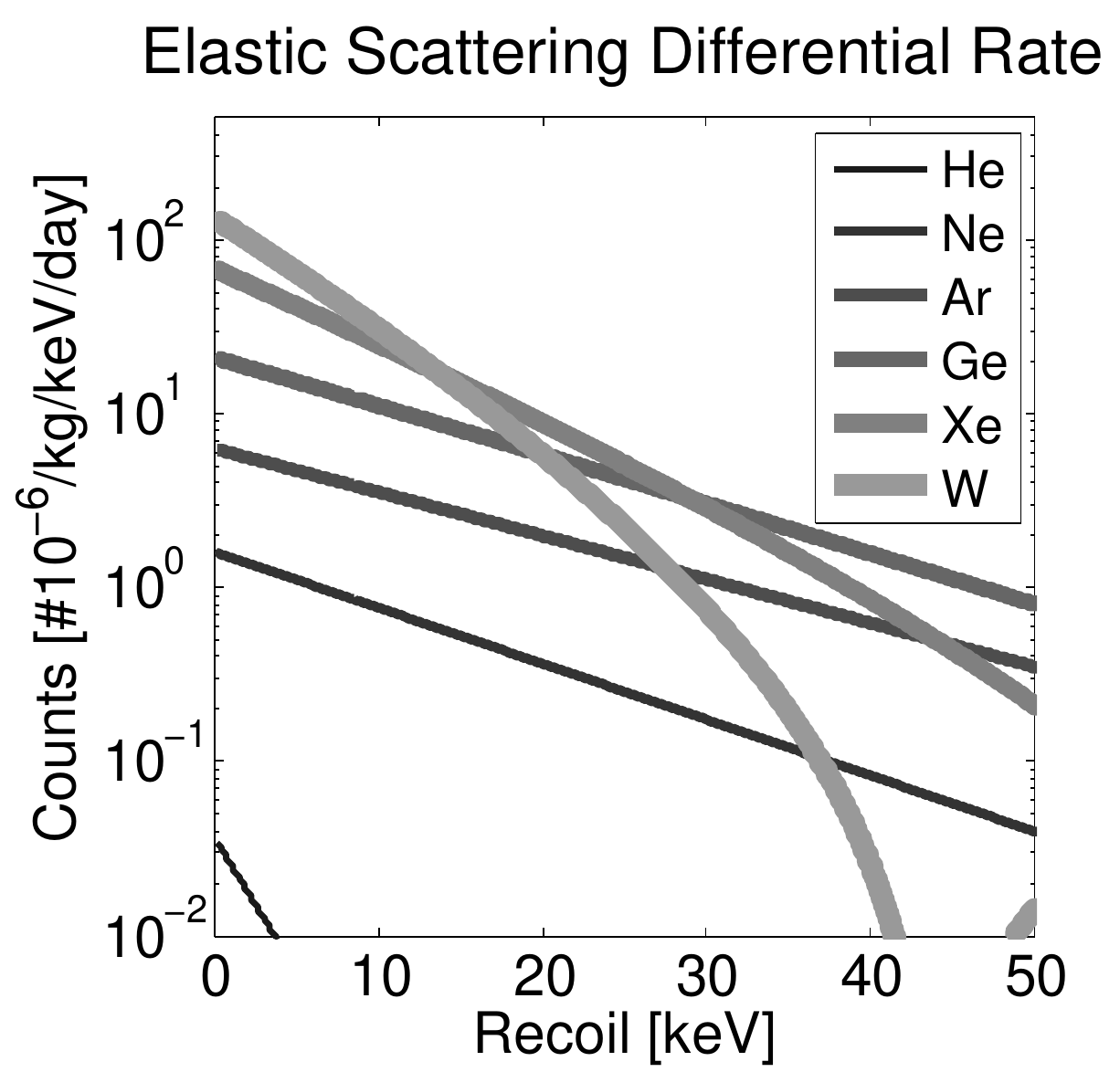}
\label{fig:1a}
}
\subfigure[][]{
\includegraphics[width=0.45\linewidth, keepaspectratio]{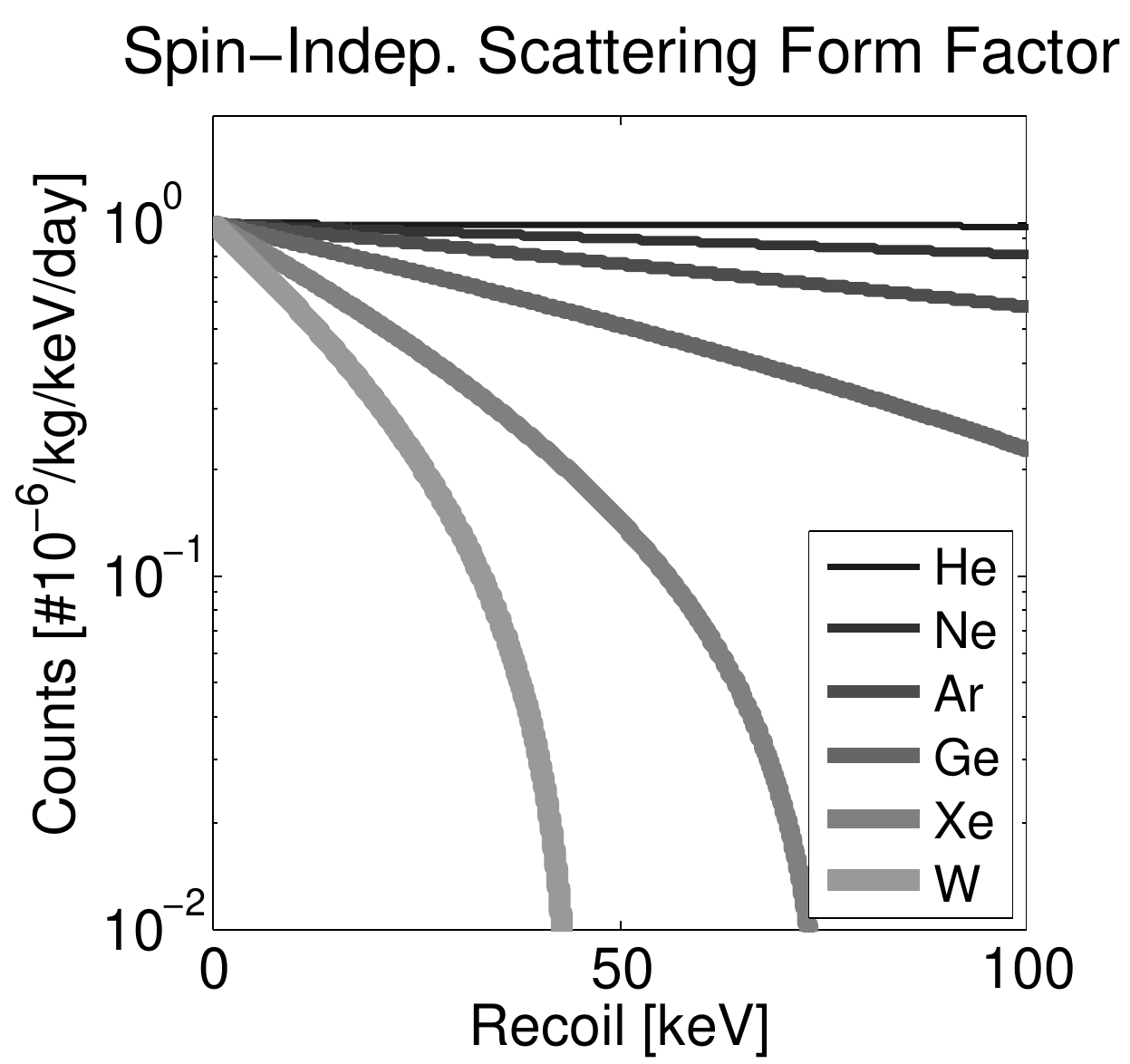}
\label{fig:1b}
}
\end{center}
\caption{Differential energy spectrum of a spin-independent elastic scattering WIMP for various nuclei. The heavier nuclei show a higher interaction rate at low recoil energies, however, they exhibit a loss of coherence for large momentum transfers.}
\label{fig:DifferentialScatteringRate}
\end{figure}\Fref{fig:DifferentialScatteringRate} shows the expected elastic scattering spectra for an arbitrarily chosen WIMP of mass ($m_\chi=100\Gcc$) and spin-independent cross-section ($\sigma_{\chi -p} = 10^{-45}\cmm$) with various target nuclei. It can be seen that the spectra are roughly falling exponentials, with an average energy of $\approx$ tens of keV. The advantage of the $A^2$ scaling of the scattering cross-section is somewhat mitigated by the form factor $F(Q)$ of the heavier nuclei, leading to a sharper fall-off at high momentum transfers. The total event rate in the detector is given by the integral of this curve with the integration bounds, in principle, being $Q_{\mathrm{min}}=0$, the lowest possible energy transfer to the recoiling nucleus, and $Q_{\mathrm{max}}=\left(\frac{2m_N}{\left(1+\frac{m_N}{m_\chi}\right)^2}\right) v_{esc}^2$, the largest possible energy transfer with $v_{esc}$ being the escape velocity of WIMPs form the galactic halo. In practice, all detectors have a lowest energy to which they are sensitive, referred to as the threshold energy $Q_{\mathrm{th}}$, and for most experiments is typically in the 5-40~keV range. The total rate is therefore the integral over $Q_{\mathrm{th}}$ to $Q_{\mathrm{max}}$ and increases rapidly with lower values of $Q_{\mathrm{th}}$.
It should be noted that $Q_{\mathrm{th}}$ represent the threshold energy imparted to the recoiling nucleus from a WIMP interaction. Many experiments use a readout channel into which only a fraction of recoil energy is deposited. In such a scenario, the measured energy has to be converted to the true recoil energy through an energy-dependent factor called the quenching parameter~\cite{Lindhard:1964vk}. The measured energy is often labeled in units of keVee (keV electron equivalent), while the nuclear recoil energy is often labeled with units of keV, keVr, or keVnr. 

For the values of $m_\chi$ and $\sigma_{\chi -p}$ assumed in the plot, the signal event rates are less than 2\,events/kg/year, orders of magnitude lower than the background rates achieved in the cleanest detector materials, typically $\le$ 1\,event/kg/keV/day. The ability to distinguish potential dark matter events from background interactions is therefore essential for to the success of a direct detection experiment. 


\section{WIMP Interactions in a Detector}\label{sec:BackgroundRejection}

In the energy range of interest for WIMP detection $(0-100\keV)$ the main contributions to the background event rate are electromagnetic interactions from $\alpha$-particles, electrons, and photons. These backgrounds originate from naturally occurring radioactive isotopes in the material surrounding the detectors, in airborne contaminants that can plate out on the experiments' surfaces, or from within the detectors themselves. Neutrons from natural radioactivity (both fission and ($\alpha$, n) reactions) will provide a nuclear interaction background. Appropriate shielding with passive and/or active materials as well as veto detectors around the experiment can significantly suppress the background event rate. Water, polyethylene, and other hydrogen-rich materials are well suited for moderating the neutron background whereas lead, copper, and other high-atomic-mass materials are better suited for moderating the $\gamma$ backgrounds. Appropriate selection and preparation of the raw materials for the detectors and their immediate support structures, in combination with adequately clean fabrication and operation procedures, can minimize the backgrounds from within the experiment itself. 

\begin{figure}[!ht]
   \centering
   \includegraphics[width=0.8\linewidth, keepaspectratio]{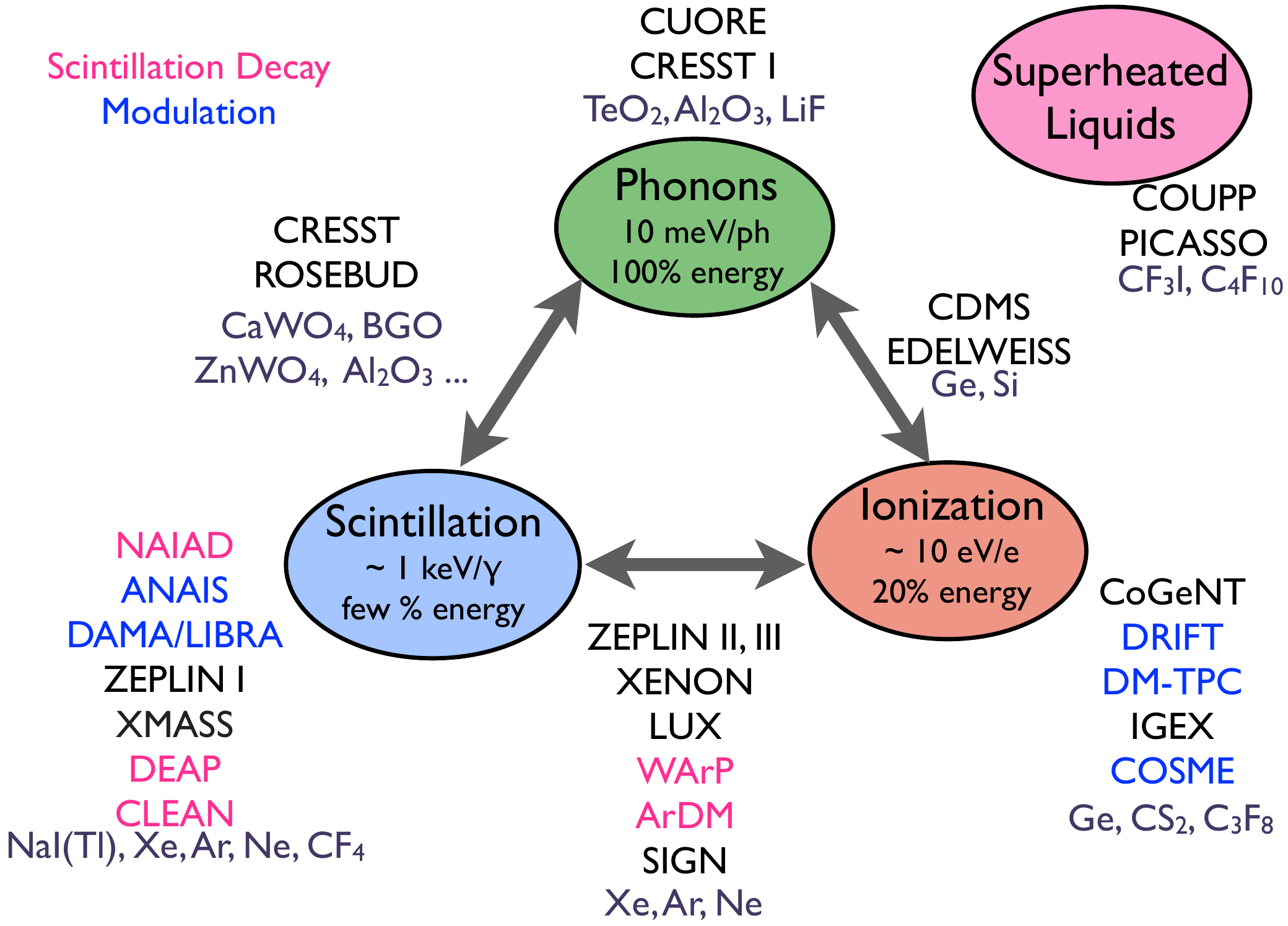}
   \caption{The corners of the triangle correspond to common energy readout channels. Experiments are listed near the main readout channel, or between the two channels used for discrimination.}
   \label{fig:tri}
\end{figure}
Various physical phenomena can be exploited to discriminate between the signal and the remaining background events. The discrimination is mostly based on the statistical properties of a large population of signal (and background) interactions, however, there also exist some techniques which permit the identification of the interaction type on an event by event basis.
\begin{enumerate}
\item Temporal variation in the signal and background event rates, generally referred to as the seasonal or annual modulation effect. This method of discrimination is based on the variation in the WIMP event rates and spectrum as the relative motion between the laboratory frame of reference and the WIMP rest frame varies along the earth's orbit of the sun~\cite{1988PhRvD..37.1353S}, whereas background sources are not expected to exhibit such a variation. 
Under the assumption of a non-rotating WIMP halo the event rate is expected to exhibit maxima/minima in June/December, with an amplitude of a few percent. Since the amplitude of the modulation is small in comparison to the overall rate, this method lends itself to experiments with large exposures and overall interaction rates.

\item Directional variation of the recoiling nucleus. 
Under the assumption of a non-rotating WIMP halo, the motion of the solar system through the galaxy results in a net WIMP wind from the direction of the solar system's motion of similar magnitude to the velocity dispersion of the WIMPs at the Sun's position in the halo.
The strong correlation between the direction of the incoming WIMP and the recoiling nucleus means that the majority of signal events, as seen from the laboratory frame, should point in the direction of the WIMP wind~\cite{1988PhRvD..37.1353S}. Background events are not expected to exhibit a non-uniform directionality, or none that is correlated with the relative directions of the laboratory frame and the WIMP wind, which varies on a 24 hour time scale as the earth rotates on its axis. Employing such a discrimination technique requires, by necessity, detectors capable of reconstructing the tracks of individual nuclei.

\item Variation in detector based response to signal and background events. In the keV momentum transfer range, signal events are due to WIMP interactions with atomic nuclei (referred to as nuclear-recoils),  whereas the majority of the backgrounds is due to electromagnetic interactions with the atomic electrons (referred to as electron-recoils).  For a given amount of imparted energy, a recoiling nucleus travels a much smaller distance in the detector than a recoiling electron. Therefore, an interacting WIMP results in a larger local deposited energy density than that of a background interaction, which can lead to a variation in the overall detector response. Detection techniques that are sensitive to this effect allow for the identification and rejection of electromagnetic background interactions on an event to event basis. For a given deposited energy, the relative scintillation, ionization, or phonon signals (or a combination thereof) are often different for electron and nuclear recoils and therefore allow for a discrimination ability~\cite{Lindhard:1964vk}. Background events resulting in nuclear-recoils, primarily due to neutron interactions within the detector, however cannot be identified and rejected with such a technique. \Fref{fig:tri} shows some of the  experiments which make use of either the statistical or event-by-event discrimination strategies along with the readout mechanism employed.

\item Operation at a deep underground site to reduce the flux of energetic neutrons. 
Regardless of the experimental technique, reducing the rate of background events is essential for all dark matter experiments. Energetic neutrons, which are created by the interaction of cosmic-ray muons within the materials and structures surrounding the experiment, lead to interactions in a detector that are indistinguishable from a WIMP signal. 

A large overburden of rock results in the attenuation of the muon flux, and subsequently the neutron background rates.  Since the nature of the overburden above any given laboratory varies, the depths are typically quoted in a normalized unit based on the equivalent depth of a water column. This is referred to as ``meters water equivalent (m.w.e.)".
Fig.~\ref{fig:Muons} shows the relative muon and neutron fluxes at a selection of underground labs.
\begin{figure}
	\centering
	\includegraphics[width=0.5\linewidth, keepaspectratio]{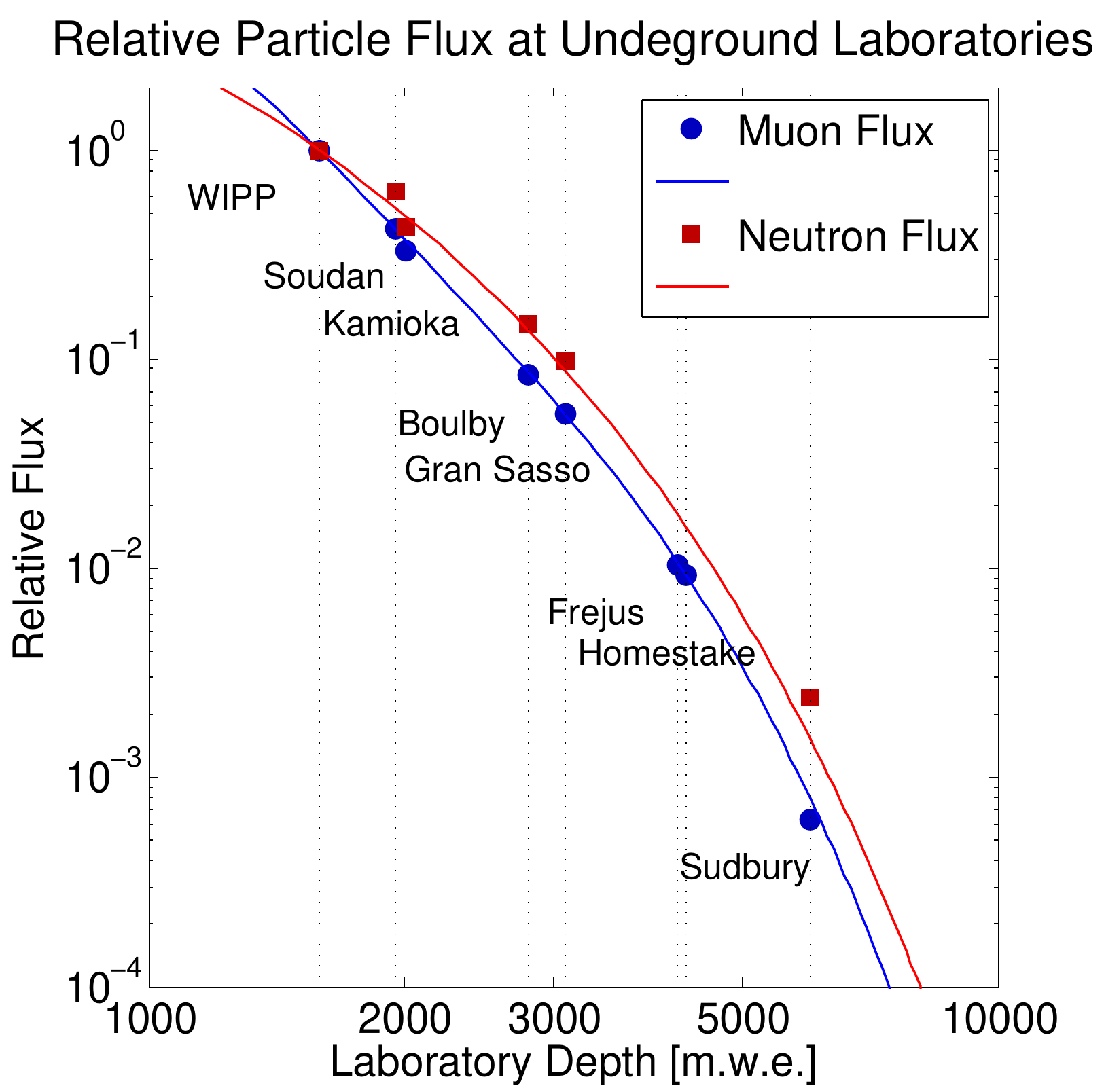}
	\caption{Normalized flux of cosmic ray muons (blue) and the resulting energetic neutrons (red) at various underground laboratories~\protect\cite{Mei:2006fr}. The blue squares are measured values, and lines is a functional fit. The neutron data are based on a Monte-Carlo simulation: the line is the total flux, whereas the circles are the flux above 10\,MeV.}
	\label{fig:Muons}
\end{figure}
\end{enumerate}


\subsection{Experimental Sensitivity}
Given the background rates and discrimination methods specific to a particular experiment, the ultimate sensitivity (in terms of the lowest WIMP scattering cross section that can be probed) of that experiment can be determined.
The four parameters required for performing this calculation are:
\begin{itemize}
\item The background rate: $B$. 
\item The background misidentification fraction: $\beta$ (i.e. the fraction of background events which may appear as signal candidates after the experimental rejection techniques are applied).
\item The signal acceptance fraction: $\alpha$ (i.e. the fraction of signal events that remain after the experimental rejection techniques are applied).
\item The exposure: $MT$ (a measure of the expected total number of signal events, where $M$ is the mass of the detectors and $T$ is the duration of the data acquisition).
\end{itemize}

\subsubsection{No background discrimination}
For experiments which do not distinguish between signal and background $\beta=1$ (i.e. all interactions within the detector cannot be individually distinguished from signal events). $\alpha$ can take any value  between 0 and 1.  

In the case of an experiment in which there were \emph{zero} observed events then an upper limit on the interaction cross-section can be determined with $90\%$ confidence from \eref{eq:zeroevents}:
\begin{equation}
S_{90} = \frac{2.3}{\alpha MT}
\label{eq:zeroevents}
\end{equation}
It can be seen that in this regime the sensitivity of an experiment increases with increase value of $\alpha$, and will improve linearly with exposure (or with time for a given detector mass). After a sufficiently large exposure the experiment will begin to observe events. In the scenario where the background rate is extremely larger than the signal rate then all such events will be assumed to be background events and the interaction cross-section upper limits is then given by \eref{eq:Nbkg}:
\begin{equation}
S_{90} \propto \frac{N_{bkg}+1.28\sqrt{N_{bkg}}}{\alpha MT} \longrightarrow \frac{B}{\alpha} + \frac{1.28}{\alpha}\sqrt{\frac{\beta B}{MT}}
\label{eq:Nbkg}
\end{equation}
At that point the sensitivity will quickly reach an asymptotic value of $S_{90} \propto \frac{B}{\alpha}$. The specific backgrounds associated with a particular experiment  determine the ultimate sensitivity and in order to make any further improvements one must find a way to reduce the background rate incident on the detectors.

\subsubsection{Background discrimination}
For experiments which do distinguish between signal and background then the parameters $\beta$ and $\alpha$ take on values in the following range: $0<\beta<1$ and $0<\alpha<1$. Typically the value of $\beta$ can chosen after the data has been acquired (i.e. it is determined during the data analysis) based on an event parameter ($\eta$) which describes the detector's differing response to signal and background events. For example, if background events exhibit a shorter output pulse rise time than signal events once can chose to only accept events having a rise time larger than a specific value. 

\begin{figure}
\begin{center}
\subfigure[][]{
\includegraphics[width=0.45\linewidth, keepaspectratio]{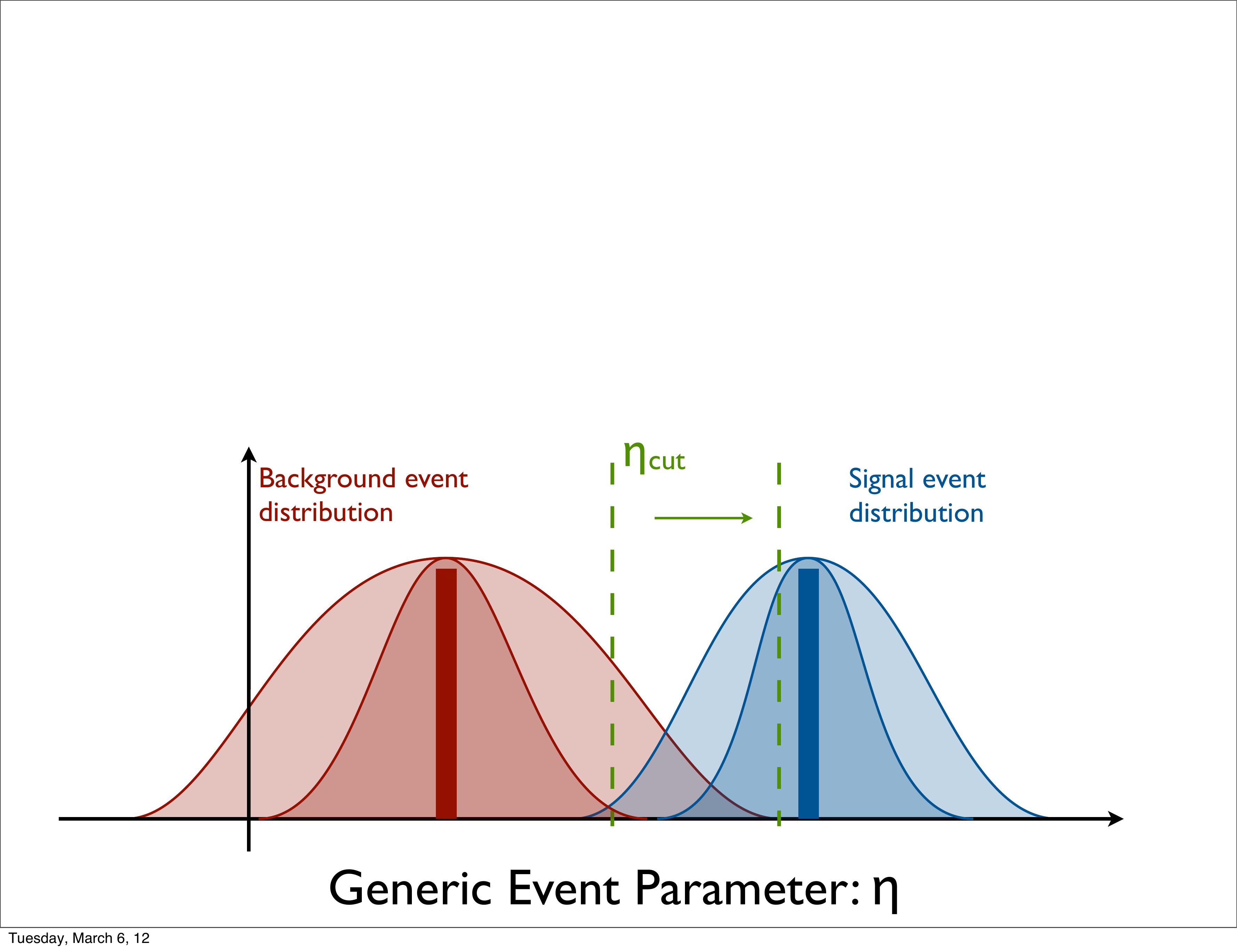}
\label{fig:discrimination}
}
\subfigure[][]{
\includegraphics[width=0.45\linewidth, keepaspectratio]{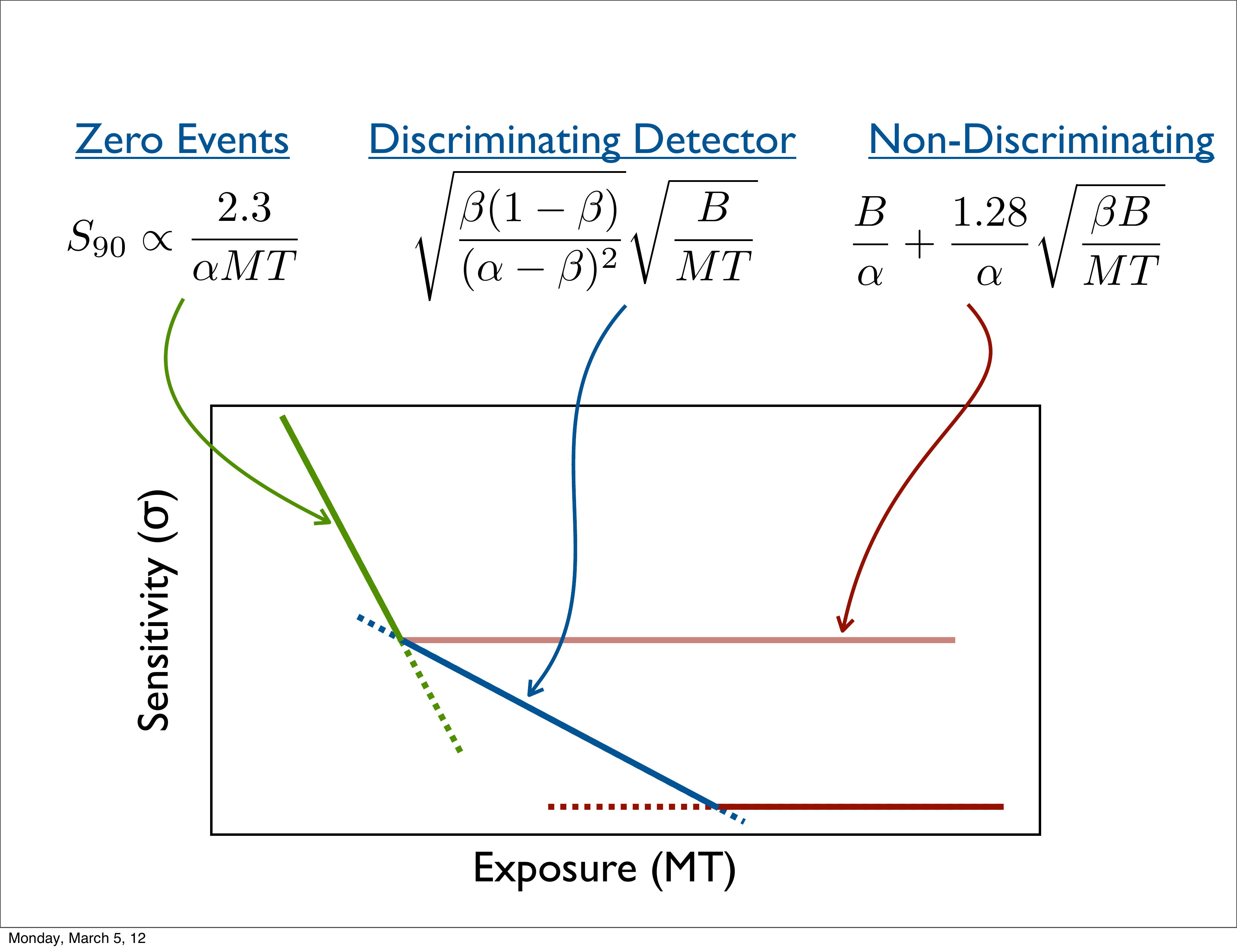}
\label{fig:sens}
}
\end{center}
\caption{(a) The distribution of the background rejection parameter ($\beta$) and signal acceptance fraction ($\alpha$) as a function of some event parameter $\eta$. The dashed lines illustrate schematically how a tighter cut ($\eta$ moved to the right) would need to be applied to maximize sensitivity as the overlap between the signal and background distributions increases. (b) Illustration of the evolution of an experiments sensitivity with increasing exposure for the cases of 1. no observed events, 2. Some observed events which are rejected via a discrimination cut, and finally 3. the ultimate sensitivity is reached.}
\label{fig:experimental-sensitivity}
\end{figure}

For a specific detector material and readout technique $\beta$ and $\alpha$ will have a characteristic distribution function (i.e. $\beta(\eta)$ and $\alpha(\eta)$) that can be determined from detector calibrations (\fref{fig:discrimination}). 
The statistical sensitivity (in the presence of background) of such an experiment is given by \eref{eq:Sstat}:
\begin{equation}
S_{Stat} \propto \sqrt{\frac{\beta(1-\beta)}{(\alpha-\beta)^2}} \sqrt{\frac{B}{MT}}
\frac{N_{bkg}+1.28\sqrt{N_{bkg}}}{\alpha MT} \longrightarrow \frac{B}{\alpha} + \frac{1.28}{\alpha}\sqrt{\frac{\beta B}{MT}}
\label{eq:Sstat}
\end{equation}
A value of $\eta$ can be chosen to minimize the quantity $\sqrt{\frac{\beta(1-\beta)}{(\alpha-\beta)^2}}$. Values as low as $10^{-3}$ can be achieved with some discrimination techniques. Choosing an overly aggressive background rejection criteria (i.e. $\beta \rightarrow 1$), however, will result in diminishing returns since the corresponding value of $\alpha$ will also be small.

As in the non-discriminating case, the sensitivity of an experiment will improve linearly with exposure until the first (background) events are observed. Subsequently, however, the discriminating experiments' sensitivity will continue to improve as the square root of exposure. The ultimate sensitivity of discriminating experiments is determined by the systematic errors associated with the values of $\beta$ and $\alpha$. This is illustrated in \fref{fig:sens}

In the remainder of this article I will review a variety of experimental techniques and implementations which are currently active in WIMP searches. This list is not meant to be exhaustive, but rather, highlights a variety of approaches in terms of detector target material and readout technology. A comprehensive survey of direct detection experiments can be found in references~\cite{Gaitskell:2004gd,CHARDIN:2005vu,Morales:592495,Mosca:2003xg}
The reader is also directed to Ref.~\cite{Bertone:2005bi} for a review of dark matter candidates.


\section{Direct Detection Techniques and Experiments}\label{sec:Experiments}

\subsection{Bulk Scintillators}

\begin{figure}[t]
   \centering
   \includegraphics[width=0.9\linewidth, keepaspectratio, trim= 0 285 0 0, clip]{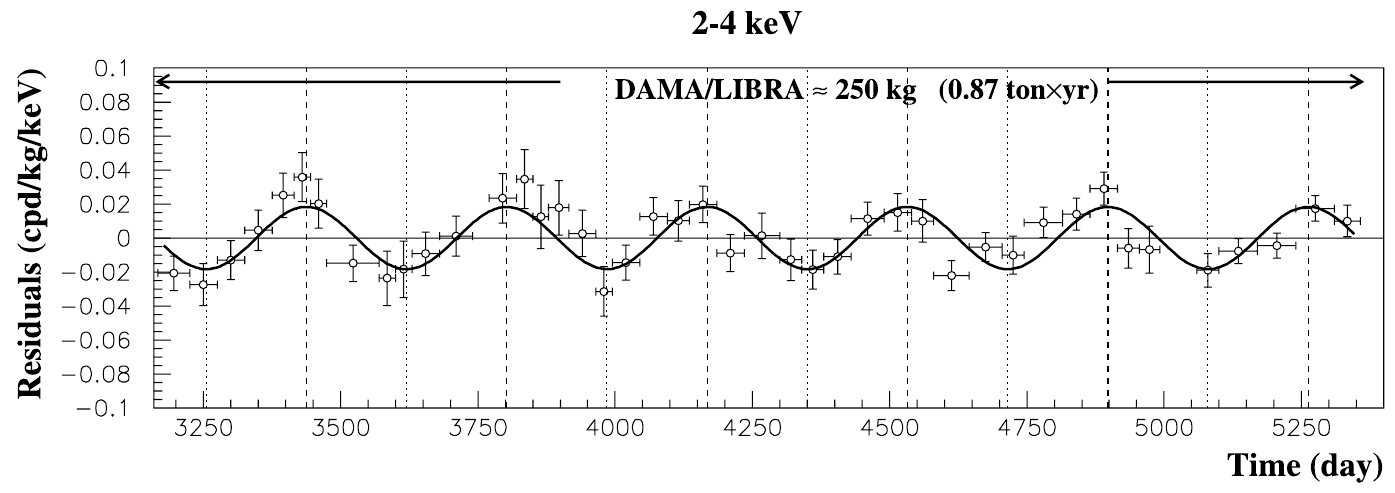}
   \caption{Experimental model-independent residual rate of the single-hit scintillation events, measured by DAMA/LIBRA, in the 2--4\keV energy interval as a function of time~\cite{2010EPJC...67...39B}. The experimental points present the errors as vertical bars and the associated time bin width as horizontal bars. The superimposed curves are cosine functions with a one year period and a maximum occurring on June 2$^\mathrm{nd}$, and with modulation amplitudes equal to the central values obtained from the best fit to the entire dataset including the exposure previously collected by the former DAMA/NaI experiment~\cite{Bernabei:2008ik}. The total cumulative exposure is 1.17 ton-yr. The dashed vertical lines correspond to the maximum and minimum expected for the DM signal (June 2nd).}
   \label{fig:dama}
\end{figure}

The DAMA/LIBRA (and formerly DAMA/NaI) is an example of an experiment which use of solid scintillators (primarily thallium-activated sodium iodide crystals, but other materials have been used) encased in a low-radioactivity enclosure and read out with photomultiplier tubes to detect particle interactions. The scintillation signal is proportional to the amount of energy imparted to a recoiling electron. For nuclear recoils (such as a potential WIMP signal) the amount of scintillation is suppressed by the quenching factor of 0.30 (0.09) for Na (Iodine) recoils~\cite{Bernabei:1996tx}. Although there is a nominal different in the pulse shape between electron and nuclear recoils this effect is too weak to exploit on an event-by-event basis. The DAMA/LIBRA experiment, instead, uses the temporal variation effect to search for a WIMP signal in their data.

The experiment is composed of 25 highly radio-pure NaI(Tl) crystals arranged in a $5\times5$ grid. Each detector, with a mass of 9.7\,kg, is encapsulated in a pure copper housing with quartz light guides coupled to a photomultiplier tube at two opposing faced of the crystal.
The detectors are placed in a low-radioactivity sealed copper box installed in the center of a multilayer Cu/Pb/Cd-foils/polyethylene/paraffin structure for shielding. 

For a cumulative exposure of 1.17\,ton-yr covering a period of 13 annual cycles an event rate of $\sim$1~count/kg/keV/day above an energy threshold of 2\,keVee (electron-equivalent) is measured. An annual modulation of the event rate is observed that is consistent with the phase, amplitude, and spectrum expected from a non-rotating WIMP halo~\cite{2004IJMPD..13.2127B,Bernabei:2008ik,2010EPJC...67...39B}. 
The modulation signal, which has a statistical significance of 8.9~$\sigma$ implies a dark matter event rate on the order of 0.4~counts/kg/keV/day at the lowest energy bin (around 2~keVee).

Although the WIMP--nucleon cross section calculated from this data is compatible with some minimally constrained supersymmetric models (MSSM)~\cite{Bottino:2008bk}, it is incompatible with the current upper--limits from other experiments~\cite{Ahmed:2011gy} (see Fig.~\ref{fig:limits}). A number of possible explanations for this discrepancy have been proposed, ranging from non-MSSM dark matter candidates and variations in the dark matter halo model to unaccounted physical effects in the detectors and concerns over issues with the data and/or its interpretation. For the time being, however, non of the proposed explanations offers a satisfactory resolution to the incompatibility of the various experimental results.

Other experiments using NaI crystals include ANAIS~\cite{Amare:2010zz}, NAIAD~\cite{Alner:2005kt}
 and ELEGANT~V~\cite{Yoshida:2003py} with a proposed experiment, called DM-Ice, planning to deploy NaI crystals 2~km beneath the south pole ice  surface (within the volume of the IceCube experiment) to search for the annual modulation signature at the south pole. Other crystals, such as CsI used by the KIMS experiment~\cite{2008AIPC.1078..533M}.


\subsection{Ionization Detectors}

\begin{figure}[t]
   \centering
   \includegraphics[width=0.9\linewidth]{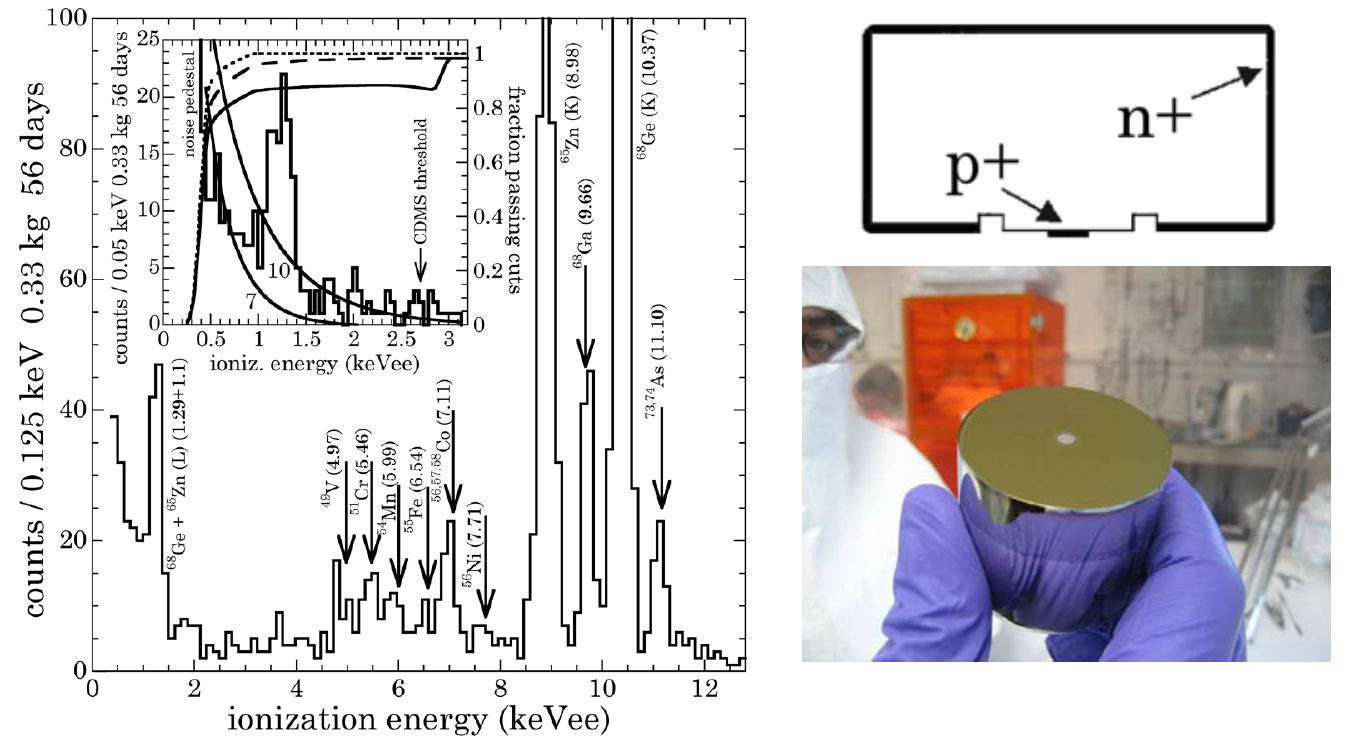}
   \caption{Left: Low energy CoGeNT spectrum after all cuts, prior to efficiency corrections. Arrows indicate expected energies for all viable cosmogenic peaks. Inset: Expanded threshold region, showing the $^{65}$Zn and $^{68}$Ge L-shell electron capture peaks. Overlapped on the spectrum are the sigmoids for triggering efficiency (dotted), trigger $+$ microphonic PSD cuts (dashed) and trigger $+$ PSD $+$ rise time cuts (solid, obtained via high-statistic electronic pulser calibrations. Also shown are the reference signals (exponentials) from 7 and 10~GeV/c$^{2}$ WIMPs with spin-independent coupling $\sigma_\mathrm{SI}=10^{-40}$~cm$^{2}$. Right: Schematic and picture of CoGeNT PPC detector.}
   \label{fig:cogentspec}
\end{figure}
\begin{figure}[t]
   \centering
   \includegraphics[width=0.75\linewidth, keepaspectratio, trim= 0 235 0 0, clip]{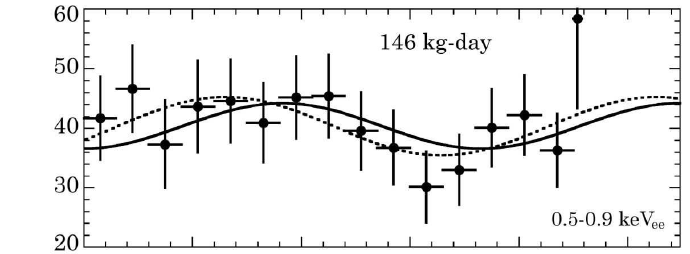}
   \caption{Evidence of modulation in the low energy spectrum of the CoGeNT detector.}
   \label{fig:cogentannmod}
\end{figure}
The CoGeNT experiment uses the ionization signal from high-purity, low-radioactivity germanium as their particle detection mechanism. The detector design called a p-type point contact (PPC) detector uses a 440~g crystal with an electrode design optimized for low capacitance in order to achieve a sufficiently high energy resolution to allow for a 0.4~keVee threshold.

Pulse shape discrimination is used to remove events due to microphonics, partial charge collection, or surface interactions. The experiment uses a combination of passive shielding (5\,cm of low-background lead, 15\,cm of standard lead, 0.5\,cm of borated neutron absorber, and 30\,cm of polyethylene) to reduce the flux of events incident on the detector and active shielding (a NaI anti-Compton veto and muon veto) to reject events which are not fully contained within or originate outside of the detectors. An exposure of 18.48\,kg-day yielded the data in \fref{fig:cogentspec}. The spectrum is dominated by known electron-capture and cosmogenic activation lines from Ge and impurities in the material. The inset in the figure shows the low-energy band, from 0.4 to 3.2~keVee, which was used for their WIMP search. As shown in the inset, there is a not-accounted-for monotonically decreasing feature in the spectrum at the lowest energies. 
Although this was originally interpreted as a signal from a light-mass WIMP~\cite{2011PhRvL.106m1301A}, the majority of those were subsequently identified as residual surface events, significantly weakening the claim of a dark matter signal.

The Cogent data was also analyzed for evidence of an annual modulation in the event rate. The obtained modulation is shown in \fref{fig:cogentannmod}. A best fit to this modulation produces a minimum modulation phase of October 16th (compared with December 2nd for DAMA/LIBRA) and a modulation amplitude of 16\%. It should be noted, however, that the date is consistent with zero modulation at the 16\% confidence level~\cite{Aalseth:1356373}.

The dark matter interpretation of the CoGeNT signal was been excluded by the XENON~100 collaboration~\cite{2010PhRvL.105m1302A}, however, the validity of this comparison is based on the correct extrapolation of the quenching factor in Xe to low energies, which is a topic of active debate~\cite{Collar:2010tm,2010arXiv1005.2615T,Collar:2010gd,2010JCAP...09..033S}

The CDMS experiment (discussed in Section~\ref{sec:cdms}) measures the phonon signal in their detector, which allows a precise measurement of the recoil energy. Their germanium detector's energy scale is well calibrated and is minimally affected by quenching factor arguments for electron recoils. A recent analysis of the CDMS-II data conclusively rules out the interpretation of both the DAMA/LIBRA and CoGeNT signals as spin-independent elastic scattering from a standard isothermal galactic halo of weakly interacting massive particles~\cite{Ahmed:2011gy}.

IGEX~\cite{Morales:524321} and TEXONO~\cite{Lin:1074769} are among the list of experiments which also make use of Ge ionization detectors for dark matter searches.


\subsection{Threshold Detectors}

\begin{figure}
\begin{center}
\includegraphics[width=0.85\linewidth, keepaspectratio]{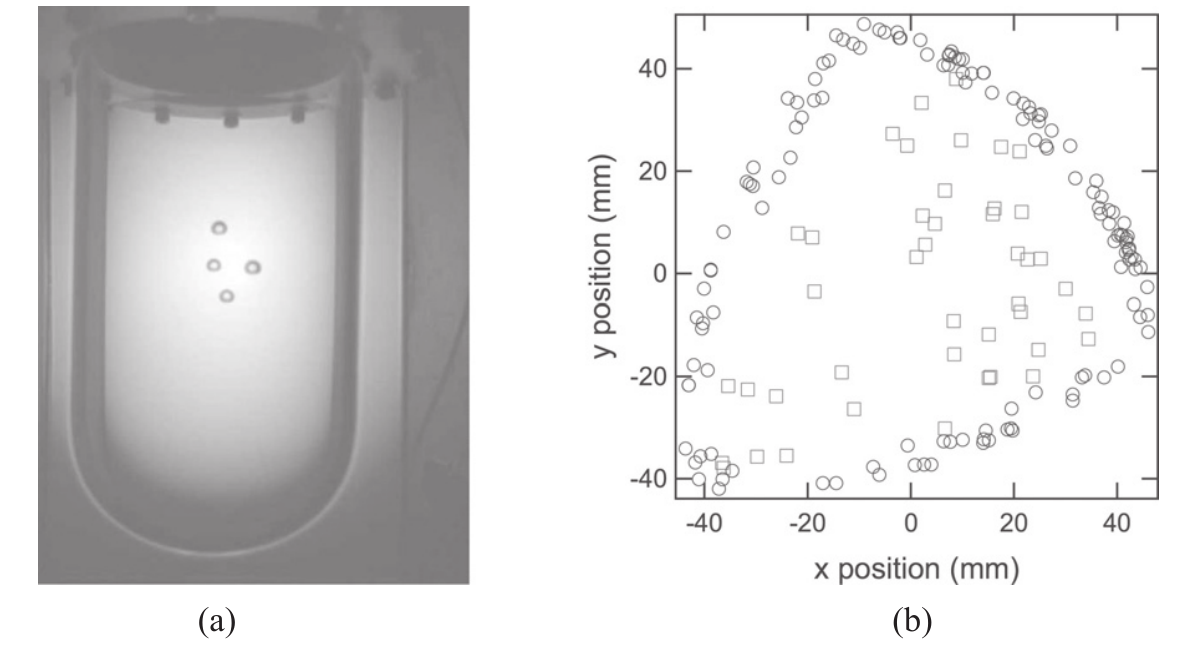}
\end{center}
\caption{(a) Photograph of the COUPP detector showing a multiple-scatter neutron event. (b)
Reconstructed event position in the x-y plane, showing a distinction between events at the surface of the vessel and ones in the bulk. Figure reproduced from Ref.~\protect\cite{Bolte:2006fq}.}
\label{fig:COUPP_Det}
\end{figure}

The COUPP (Chicagoland Observatory for Underground Particle Physics) experiment makes use of a bubble chamber detector to search for dark matter~\cite{Behnke:2008tn}. The bubble chamber contains a superheated liquid (CF$_{3}$I) which is the target material. An energy deposition in the liquid from a particle interaction leads to a local nucleation of a bubble at the interaction site. The pressure spike due to the bubble formation trigger a imaging sensors (video cameras) placed around the detector to record the the event, allowing for the 3-dimensional position reconstruction of the interaction site as shown in \fref{fig:COUPP_Det}.

\begin{figure}
\begin{center}
\subfigure[][]{
\includegraphics[width=0.45\linewidth, keepaspectratio]{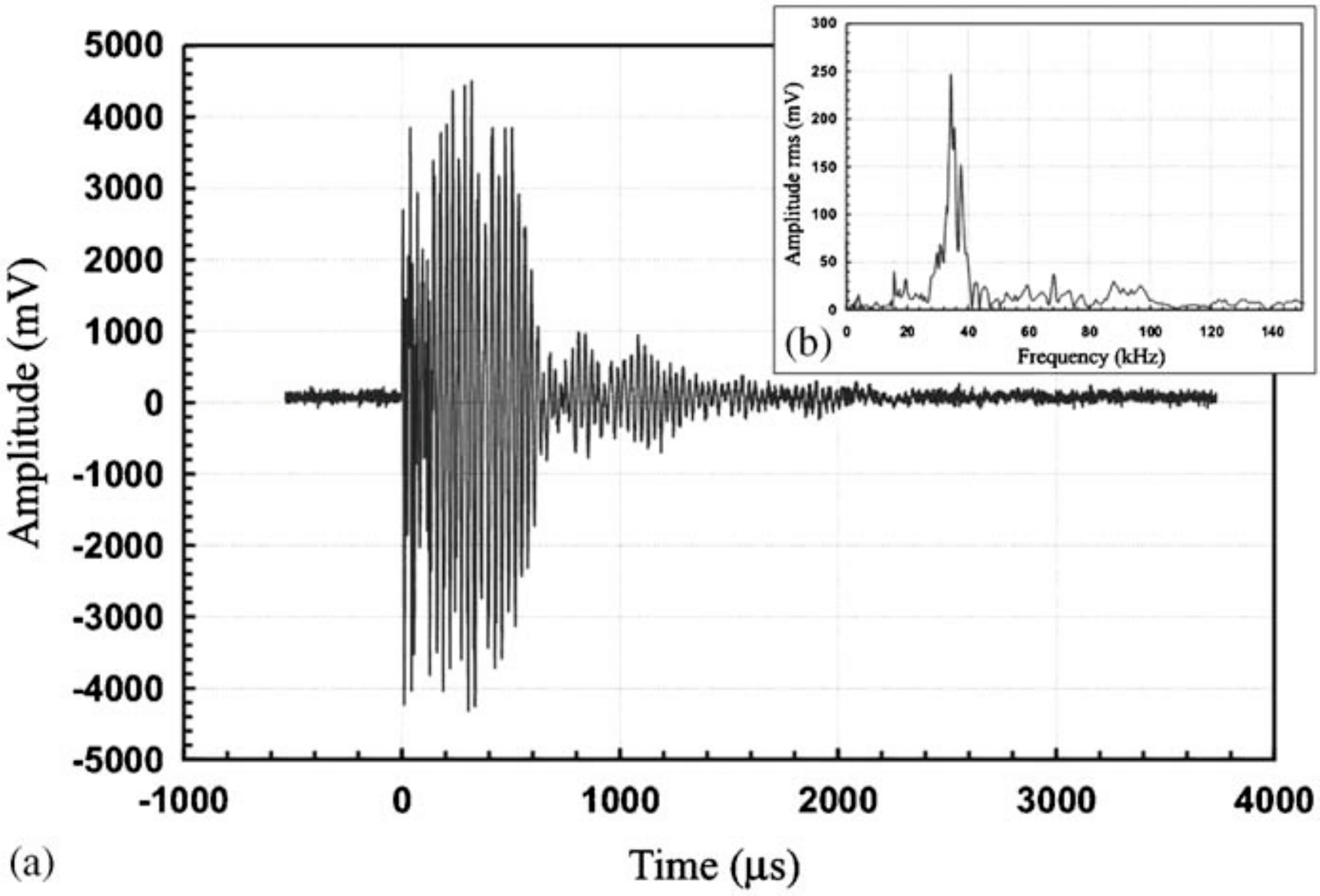}
\label{fig:PICASSO_Signal}
}
\subfigure[][]{
\includegraphics[width=0.4\linewidth, keepaspectratio]{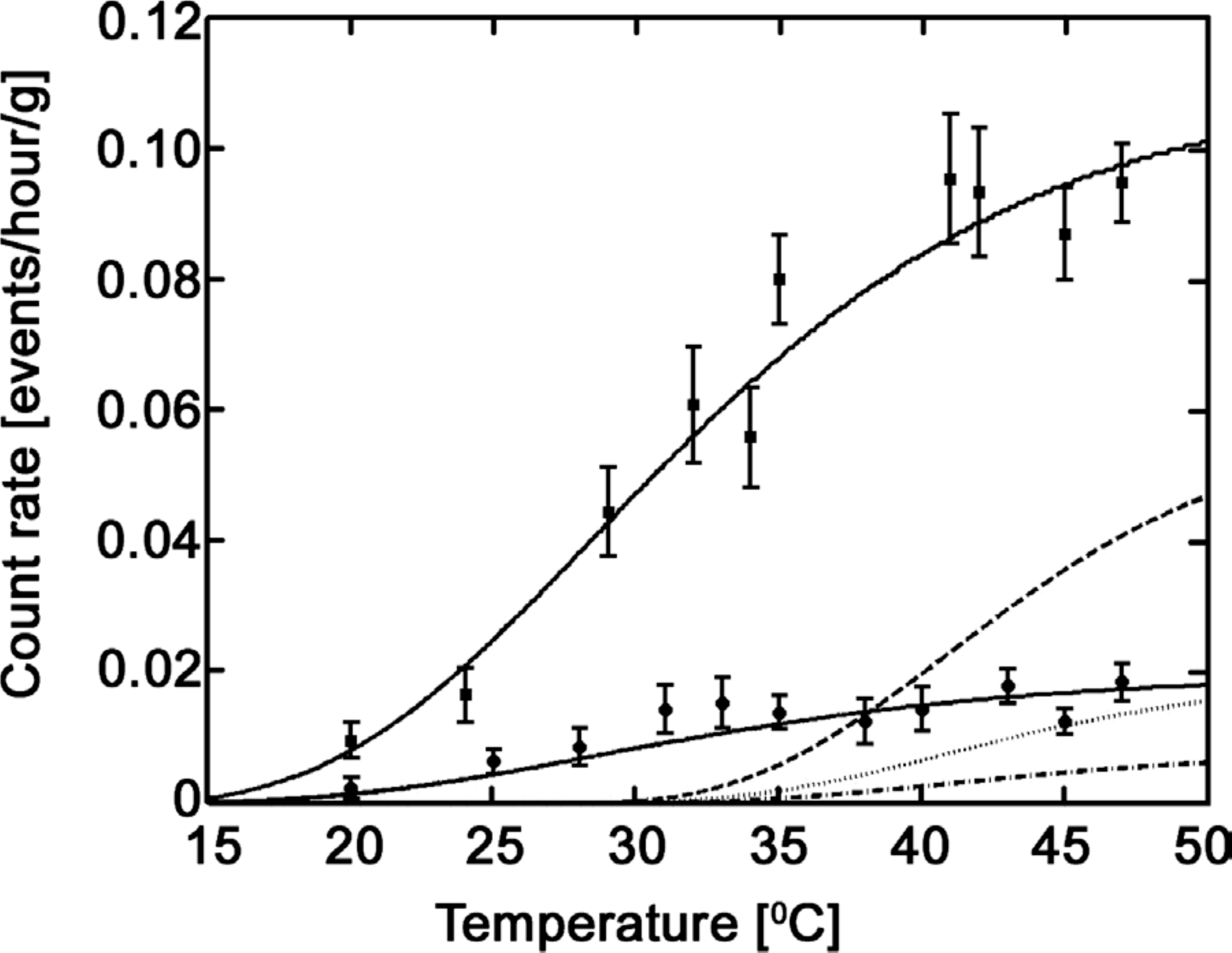}
\label{fig:PICASSO_Rates}
}
\end{center}
\caption{(a) Acoustic signal of an event in the PICASSO detectors. The inset show the Fourier transform of the signal. Figure reproduced from Ref~\protect\cite{Barnabe-Heider:876803}. (b) Measured event rates (points) as a function of detector temperatures. Also shown are the expected event rates for $\alpha$ particles (solid lines) and a 50 GeV/c$^2$ WIMP signal. Figure reproduced from~\protect\cite{Barnabe-Heider:822461}.}
\label{fig:PICASSO}
\end{figure}

An important property of the bubble chamber is that the energy loss per unit of path distance ($dE/dx$) required to nucleate a bubble depends strongly on the temperature and pressure of the superheated liquid. Therefore, by carefully controlling the operating conditions, interactions with very high $dE/dx$ such those from nuclear recoils will result in bubble formation while electron recoil events will not. Thus, the detector is inherently insensitive to the largest background sources. The bubble formation process does nor allow the determination of the interaction energy. The experiment operates in a mode of counting the total integral interaction rate above a threshold energy. The threshold energy, however, is sensitive to the operating temperature and pressure, therefore it is possible to statistically determine an energy distribution of the interaction by taking data at different thresholds (\fref{fig:PICASSO_Rates}). A 60~kg detector is currently being commissioned at Fermi National Laboratory and will be installed at a deep site in 2011.

The Picasso experiment works on a similar principle to COUPP. The detector is made of superheated C$_{4}$F$_{10}$) droplets which are embedded in a gel matrix. The detectors are assembled into 4.5~liter modules each containing 85~g of target material. Acoustic sensors detect the occurrence of an interaction, and the frequency spectrum of the recorded signal can be used to discriminate between events caused by low energy nuclear recoils and $\alpha$-particle interactions (\fref{fig:PICASSO_Signal})~\cite{Archambault:2009bx}.


\subsection{Directional Detectors}

Gaseous detectors allow the possibility of reconstructing the track of a recoiling nucleus. As mentioned in \sref{sec:BackgroundRejection} this information is very useful in distinguishing a dark matter signal from among the ever-present background by exploiting the diurnal modulation of the dark matter ``wind''. The disadvantage of this approach, however, is the relatively small target mass, and thus exposure, that can be readily achieved.

The DRIFT (Directional Recoil Identification From Tracks) experiment employs a negative on time projection chamber filled with CS$_2$ gas as a Dark Matter detector. An interaction occurring within the detector results in a recoiling nucleus and the creation of electron-ion pairs along the track. The CS$_2$ ions drift towards the anode and are read out via multi-wire proportional chambers, which yield the position of the track in two dimension, as well as the energy of the inter- action ,which is proportional to the total number of ions. The third dimension of the recoil track is determined from the ion drift time. DRIFT is able to do discriminate between different classes of events, such as: electron-recoils, nuclear-recoils, and $\alpha$ particles, based on the track morphology. In this setup the direction of the recoiling nucleus is determined up to an overall $180^\circ$, i.e. the starting and ending points of the track cannot be identified. 

The ability to break this degeneracy (referred to as the ``head tail effect'') provides a stronger handle in extracting a dark matter signal form the background. The DM-TPC experiment~\cite{2010APS..APRA10001B} has demonstrated this capability in a low pressure CF$_{4}$ 10~liter detector for recoiling nuclei with energy $\gtrsim$ 50~keV.


\subsection{Single-phase Noble Liquids}

Noble liquid dark matter detectors are an interesting and promising class of detector technology. In addition to being capable of being operated in single readout mode as a scintillation only detector (similar to the solid scintillator detectors) they can also be operated in a manner that makes possible a strong event by event rejection of electron recoils. The advantages of working with noble liquids include:
\begin{itemize}
\item The ability to suppress internal background sources by continuously circulating and purifying the liquids.
\item The ability to suppress external background sources via shielding the fiducial volume with active detector material
\item The ability to scale to large mass detectors. Simply put, an increase in the linear dimensions of the liquid container leads to a cubic increase in the detector mass of the detector material.
\end{itemize}

\begin{figure}
   \centering
\includegraphics[width=0.75\linewidth, keepaspectratio]{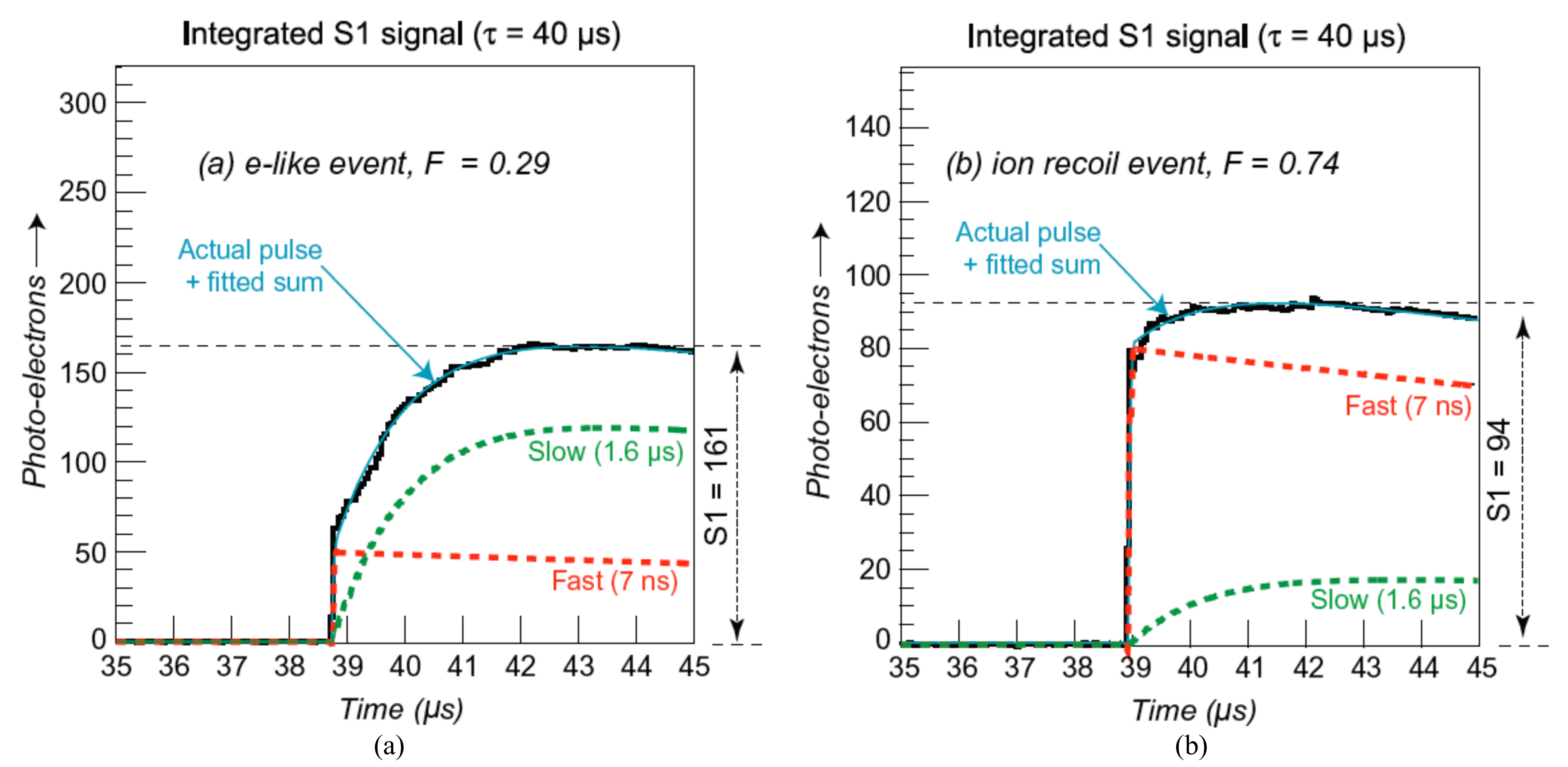}
\caption{Leading edge of the photomultiplier pulse for an electron-recoil (a) and a nuclear-recoil (b) event in liquid Argon showing the differing fast and slow pulse components. Figure reproduced from Ref.~\protect\cite{2008APh....28..495W}.}
\label{fig:WARP_Risetime}
\end{figure}

Noble liquid experiments share the same underlying detection mechanism. An interaction in the liquid results in the ionization and  excitation of the target atoms. The process through which the excited atoms go through as they decay to the ground state involves the formation of excimer states which can occur in either of a singlet or triplet state, each of which has a different decay time: 7\,ns/15.4\,$\mu$s for Ne, 7\,ns/1.5\,$\mu$s for Ar, 3/27\,ns for Xe. The fraction of singlet to triplet excimers created is different for electron and nuclear-recoils, for example, in liquid Argon 70\% of the excimers created by a nuclear-recoil are singlets, whereas the ratio is $\sim$30\% for electron-recoils~\cite{2006AcPPB..37.1997S,1983PhRvB..27.5279H}. The resulting pulse shape from a photomultiplier tube exhibits a time structure which is dependent of the nature of the recoil. \Fref{fig:WARP_Risetime} shows the difference in pulse rise-time for an electron- and nuclear-recoil event in liquid Argon.

The DEAP/CLEAN collaboration (Dark matter Experiment using Argon Pulse shape discrimination / Cryogenic Low Energy Astrophysics with Noble liquids) is building a liquid argon single-phase detector called miniCLEAN~\cite{Giuliani:2010zza}. This design is based on a spherical vessel filled with argon with all $4\pi$ steradians instrumented with photomultiplier tubes. The photon hit pattern on the photomultiplier tubes permits the reconstruction of an interaction's position within the detector, and pulse shape discrimination is used to reject electron recoil backgrounds. The miniCLEAN experiment instrument 500~kg or target material with 91 photomultiplier tubes allowing the definition of an internal low-background 150~kg fiducial volume. Within an energy window of 5--45~keVee the miniCLEAN experiment expects to achieve an exposure of 300~kg-year within two years of operation. An electron recoil rejection efficiency of $\sim 10^{-9}$ is needed to suppress the expected event rates from $\beta$ interactions. This is rejection performance is expected to be achieved with a signal acceptance fraction of $50\%$.

Similarly, the XMASS experiment is commissioning an 857~kg spherical single-phase liquid xenon detector surrounded by 642 photomultiplier tubes allowing the definition of an internal low-background 100\,kg (80\,cm diameter) fiducial volume~\cite{Sekiya:2010bf}. XMASS does not plan to make use of the pulse shape discrimination properties of the scintillation signal (which is not as pronounced as it is for argon), but rather uses the outer layer of $\sim$750\,kg of xenon surrounding the inner fiducial volume as both a passive shield and an active veto for background events originating outside of the detector allowing a dark matter search with a 5\,keVee threshold.
The XMASS detector is placed within a 10\,m diameter by 10\,m high water tank instrumented with 72 20\,inch-diameter photomultiplier tubes which acts as both a passive shield and an active veto for external neutron backgrounds.


\subsection{Two-phase Noble Liquids}

\begin{figure}
   \centering
   \includegraphics[width=0.75\linewidth, keepaspectratio]{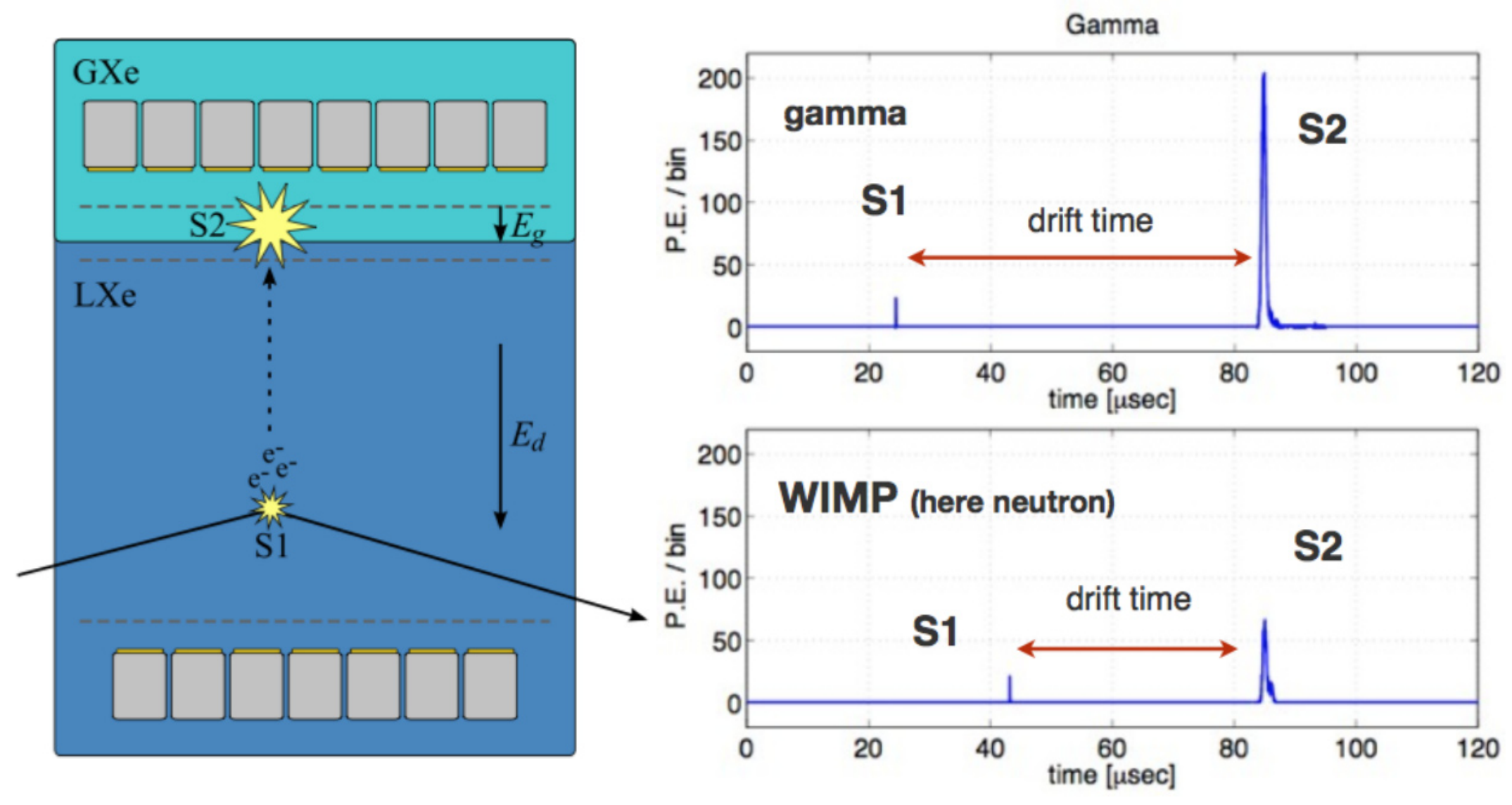}
   \caption{Schematic of a xenon two-phase time-projection chamber, showing the recorded signals for a gamma interaction and a ``WIMP'' (actually a neutron used for calibration). The ratio S2/S1 is used to discriminate between electron and nuclear recoils. Full description in the text.}
   \label{fig:xeschem}
\end{figure}

Xenon and argon two-phase time projection chambers (TPC) have also been developed for dark matter detection. A review on xenon as a detector material (with some discussion of argon as well) can be found in~\cite{2010RvMP...82.2053A}.  All experiments have a common design/operation principle which is shown schematically in \fref{fig:xeschem}.
A low-radioactivity vessel is partially filled with liquid xenon (argon), with the rest of the vessel containing xenon (argon) gas. A linear electric field of $\sim$1\,keV/cm and $\sim$10\,keV/cm are established across the liquid and gas volumes respectively by applying a voltage bias to the electrodes (shown as dashed lines in the figure).

Photomultiplier tube arrays are the top (in the gas volume) and bottom (in the liquid) of the detector are used to detect scintillation photons from the interaction. An initial signal (S1) is due to the scintillation produced at the the interaction site. Ionized electrons, which are also created by the interaction, drift in the liquid are then extracted into the gas by the high electric field. As the electron accelerate through the gas they produce a second scintillation signal (S2). \Fref{fig:xeschem} shows the S1 and S2 signals for two events: an electron recoil caused by a $\gamma$ interaction, and a nuclear recoil caused by a neutron interaction. The ratio of the S2 and S1 signals is used as a discriminator between these types of events.

The time procession chamber configuration allows for the reconstruction of an interaction's position in the detector. The pattern of hits in the photomultiplier tube arrays determines the $(x,y)$ position, whereas the time between the S1 and S2 pulses measures the drift time of the electrons through the liquid and thus determines the $z$ position. The low energy threshold of the experiment is determined by the smaller S1 signal, and is typically set at a few photons per event. For example the XENON 10 experiment measured 2.2 photoelectron/keVee, and had a threshold of 5\,keVee. 
\begin{figure}
   \centering
   \includegraphics[width=0.75\linewidth, keepaspectratio]{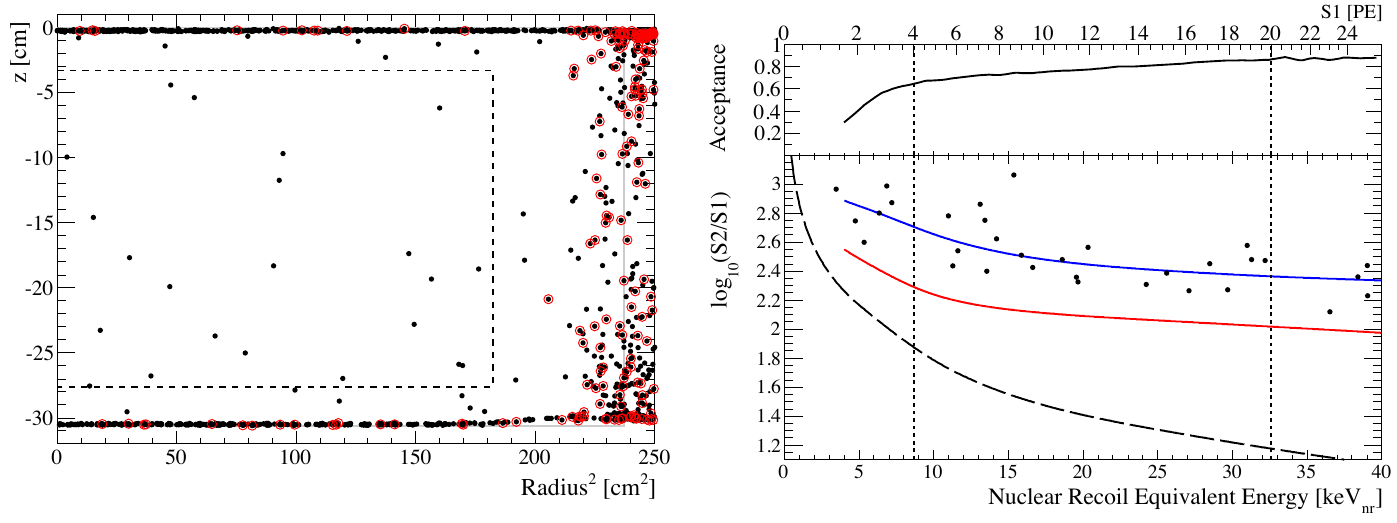}
   \caption{XENON 100 data. Left: Distribution of all events (dots) and events below the nuclear recoil median (red circles) mapped to the $(r,z)$ plane. Right: Cut acceptance (top) and $\log_{10}(S2/S1)$ (bottom) as functions of nuclear recoil energy. Only events inside the gray line region of the left plot are shown here. Colored lines correspond to the median $\log_{10}(S2/S1)$ values of the electronic (blue) and nuclear (red) recoil bands. The WIMP search window 8.7--32.6~keV$_\mathrm{nr}$ (vertical dashed lines) and S2 software threshold of 300 photoelectrons (long dashed line) are also shown. The ``WIMP  search region'' is circumscribed in this plot by the two vertical lines and the red and long dashed lines.}
   \label{fig:xe100}
\end{figure}

The WArP experiment~\cite{2011JPhCS.308a2005A} uses liquid argon as the detection material. Liquid argon has the advantage that both the S2/S1 ratio and the time constant of the light pulses have discrimination power for electron/nuclear recoils but suffers from the presence of the naturally occurring $^{39}$Ar isotope, which is a long lived $\beta$ emitter. The use of argon depleted of $^{39}$Ar both from deep reservoirs and from post-extraction processing is also being pursued.


\subsection{Cryogenic Crystal Detectors}\label{sec:cdms}

Cryogenic detection techniques offer new ways for determining the properties of an interaction in the detector, may not otherwise not be available at room temperature.
At sub-K temperatures, the following physical channels contain information relevant for direct detection experiments:
\begin{itemize}
\item The entire recoil energy of an event (independent of the nature of the recoil) can be determined from the response of the thermal and athermal phonon populations. 
\item Charge carriers can be drifted and collected with electric fields of only a few V/cm. This makes it possible to perform an ionization measurement overwhelming the signal in the phonon channel~\cite{1988JAP....64.6858L,1978JETPL..28..328N}
\item Physical effects that only exist below a certain critical temperature lead to novel readout techniques. For example:
\begin{itemize}
\item Semiconductor thermistors and superconducting transition-edge sensors allow the measurement of an event's recoil energy with excellent resolution.
\item Superconducting kinetic induction devices can be used to instrument and readout 100's to 1000's kg of detector mass.
\item Vibration wire resonators in a superfluid permit the measurement of recoil energy in a $^3$He target.
\end{itemize}
\end{itemize}

Unlike the superheated liquids, and to some extend the gaseous time projection chambers, cryogenic detectors are sensitive to all particle interactions. This means that electron-recoil backgrounds constitute the main source of the interaction rate. The ability to distinguish electron-recoils from nuclear-recoils on an event by event basis becomes paramount and permits the rejection of the great majority of such backgrounds.
A common feature among such experiments is the use of the information in the phonon (or thermal) response to determine the total recoil energy of an interaction (independent of its nature). A second detector response that is dependent on the type of recoil, such as scintillation or ionization, provides information relevant for background rejection (see \fref{fig:tri}). 

\begin{figure}
\begin{center}
\includegraphics[%
  width=0.80\linewidth,
  keepaspectratio]{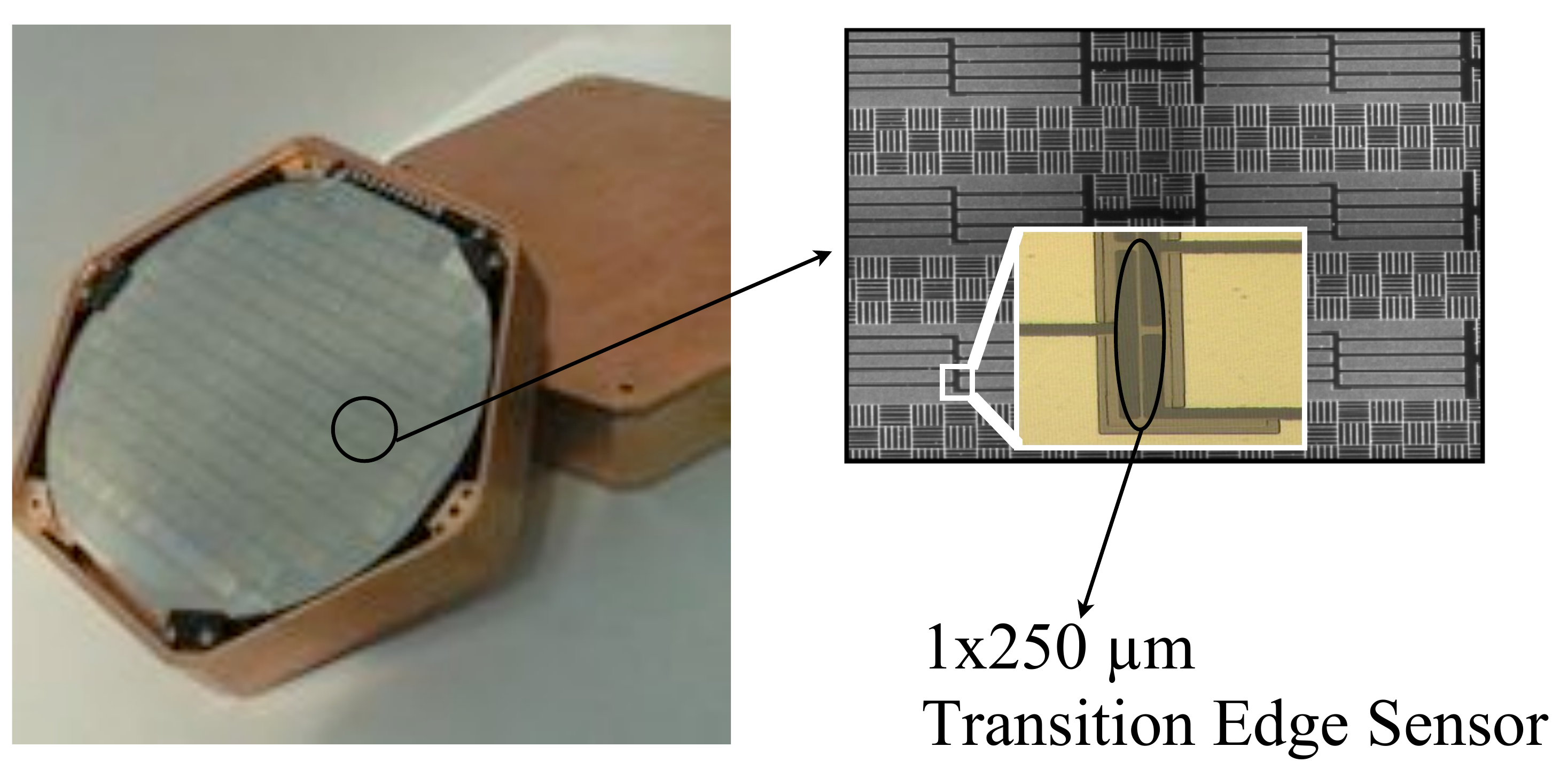}
\end{center}
\caption{A photograph of a CDMS II detector (left panel). The right panel show a zoomed in view of the TES sensor pattern on the detector surface.}
\label{fig:CDMS_Det}
\end{figure}

The Cryogenic Dark Matter Search (CDMS II) experiment uses germanium and silicon crystals which are 1~cm thick and 3~inches in diameter and as target material (\fref{fig:CDMS_Det})~\cite{Mirabolfathi:2006el,2006NIMPA.559..411A}. A fraction of the energy deposited in a semiconductor crystal by a particle interaction results in the creation of electron-hole pairs, with the remainder of the energy creating a population of high frequency (THz) athermal phonons. An electric field, applied across the crystal, causes the electrons and holes to drift to opposing electrodes, where they are subsequently measured with a charge amplifier. The high frequency phonons travel propagate through the crystal until they reach the surface where their energy is measured by Transition-Edge Sensors that are photolithographically patterned on the surface, leading to a signal that is proportional to the recoil energy~\cite{Irwin:1995vb,Irwin:1995wq}.

The measurement of the interaction energy via these two channels forms the basis of background discrimination used by the CDMS and other cryogenic based experiments. For electron-recoils, the number of charge carriers created is directly proportional to the deposited energy, whereas for nuclear-recoils, that number is suppressed roughly by a factor of three with respect to electron-recoils, and is slightly energy dependent~\cite{1992PhRvL..69.3425S}. In both cases, however, the athermal phonon channel remains sensitive to the total energy of the recoil. This variation in response allows for the identification of electron and nuclear-recoils with an efficiency of than 99.9998\%~
\cite{2004NIMPA.520..171M}. 

Events occurring within $\sim 10\,\mu$ m of the detector surfaces, have a diminished ionization response resulting in a degradation of the electron-recoil identification to 99.79\%.  The information in the time structure of the phonon pulse, however, can be used to identify event occurring near the surface versus those occurring in the bulk, and thus improve the surface electron rejection compared that based only on the ionization signal.

\begin{figure}
   \centering
   \subfigure[][]{
   \includegraphics[width=0.3\linewidth]{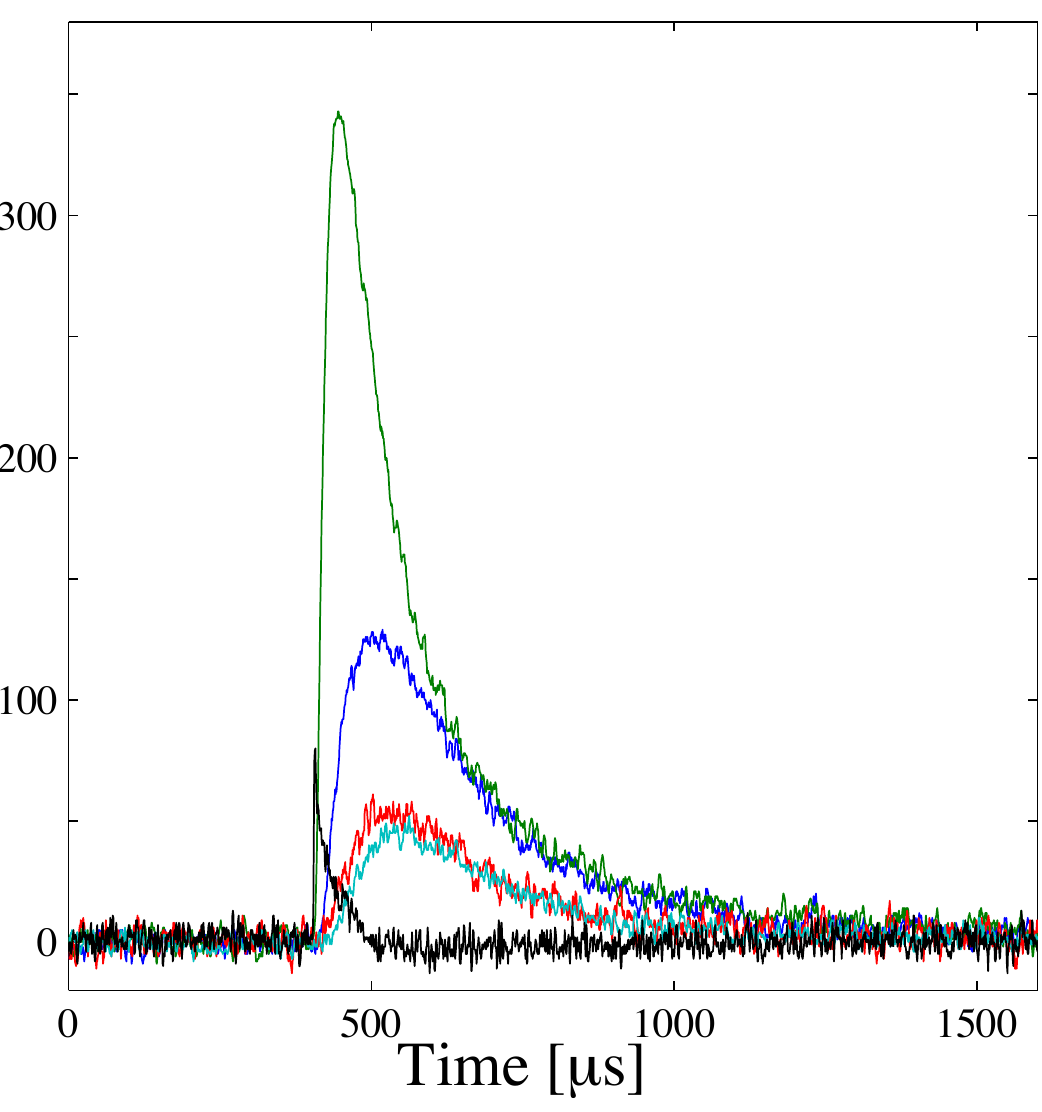}
   \label{fig:ZIP_Pulses_Ge}}
   \subfigure[][]{
   \includegraphics[width=0.3\linewidth,trim= 10 0 10 -5]{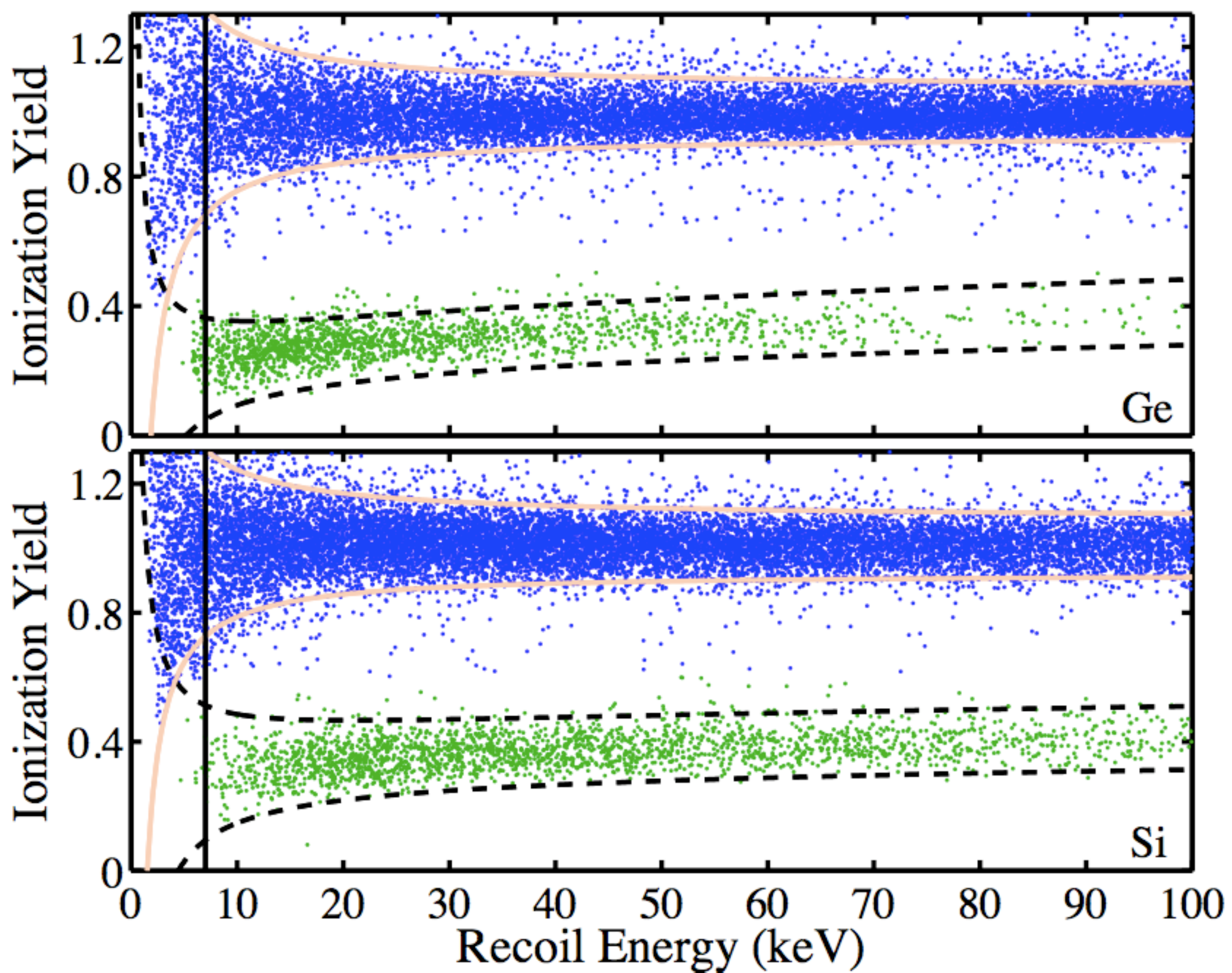}
   \label{fig:Energy_Yield}}
   \subfigure[][]{
   \includegraphics[width=0.3\linewidth,trim= 10 20 10 25]{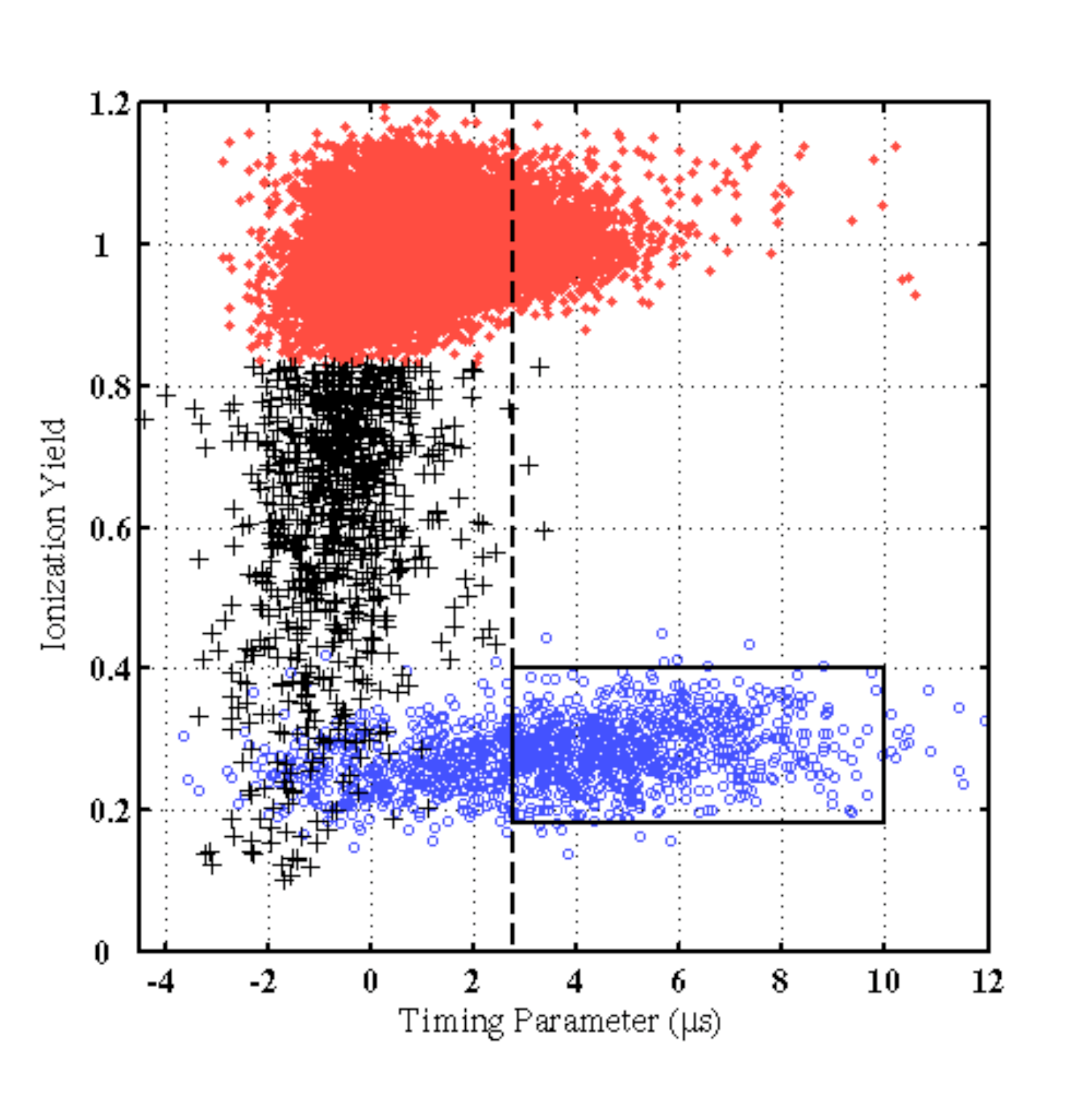} 
   \label{fig:Beta_Risetime}}
   \caption{(a) Phonon (colored) and ionization (black) pulses from a 20 keV event in a Ge detector. The variation in pulse height and timing allows the determination of an interaction'Õs position. (b) Yield vs. recoil energy for calibration data with a $\gamma$ and neutron source showing the electron (blue) and nuclear (green) recoil bands for the Ge (upper) and Si (lower) detectors. (c) Yield versus phonon-timing parameter for events in a Ge detector. The three different classes of events shown: $^{133}$Ba $\gamma$ in the bulk (red dots) and surface (black crosses) as well as nuclear-recoils (blue circles) from $^{252}$Cf neutron events, demonstrate  the background rejection capability of the detectors.}
   \label{fig:ZIP_Timing}
\end{figure}

\Fref{fig:ZIP_Pulses_Ge} shows the pulses observed for a 20 keV bulk event in a Ge ZIP. The correlation between the amplitude and rise-time of the four phonon pulses, due to a larger fraction of phonons being absorbed in the sensor nearest to the event location, permits the localization of the event in the x-y directions. The ratio of the ionization to phonon signals (defined as yield, it is equal to 1 for bulk electron-recoils, and approximately 1/3 for nuclear-recoils) is shown in Figure~\ref{fig:Energy_Yield} for electron-recoil ($\gamma$'s, blue points) and nuclear-recoil (neutrons, green points) events. Figure~\ref{fig:Beta_Risetime} shows the distribution of calibration events in yield--phonon timing parameter space~\cite{Akerib:2006dz}. Bulk nuclear-recoils (blue circles) have a larger phonon timing parameter, compared to surface electron-recoils, due to the smaller fraction of prompt ballistic phonons in the events. The clear separation between the bulk nuclear and electron-recoils allows event by event identification and rejection of bulk electron-recoil backgrounds.  It is also seen that surface electron-recoils (black crosses) cannot be rejected with an equal efficiency based on the yield parameter alone, but in combination with a phonon timing cut the rejection efficiency is greatly improved while maintaining a large acceptance of the nuclear-recoil signal.

\begin{figure}
   \centering
   \includegraphics[width=0.75\textwidth, keepaspectratio, trim=490 0 0 0, clip]{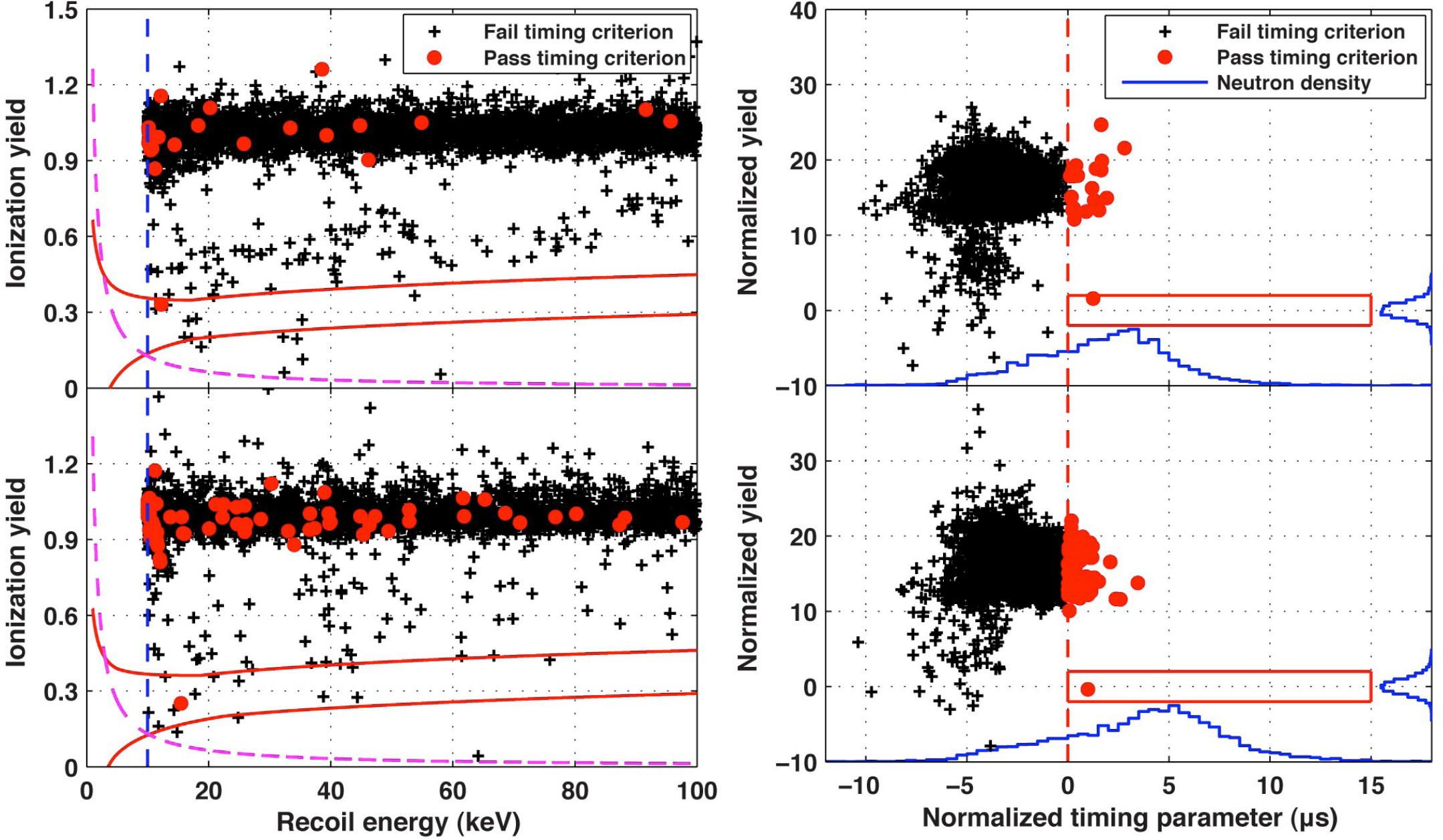}
   \caption{Left: Ionization vs. energy for the two detectors with events in the WIMP search region. A WIMP candidate must be between the red lines defining the nuclear recoil band \emph{and} pass the timing criterion (red dots). Right: Normalized yield vs. normalized timing parameter for the two detectors with events in the WIMP search region. In this figure the WIMP candidates are clear: they must fall into the red box bounded by the timing parameter on the horizontal axis and the normalized yield for the nuclear recoil band. The two events passing all cuts are seen in the box. The blue lines are the neutron calibration histograms for each detector.}\label{fig:CDMSII}
\end{figure}

\Fref{fig:CDMSII} shows the results of the last data set of the CDMS-II experiment. Two events passed all the cuts and are thus potential WIMP candidate events. Likelihood analysis done on the data set indicate it is more probable that these are leakage events and not WIMP events, but more data will need to be taken before any solid conclusions can be made. 

The CDMS collaboration is currently beginning a data run using new, more massive detectors for the SuperCDMS at Soudan experiment. The thickness of the detectors has been increased from 1~cm to 1~inch, a factor of 2.5 increase in mass per detector to 0.64~kg each. The
detectors have a new phonon and ionization sensor layout which vastly improves their surface rejection capability.
These interleaved Z-dependent Ionization and Phonon (iZIP) detectors~\cite{2009AIPC.1185..223P}, which measure charge and phonon signals on both sides of the detector, have exhibited extremely powerful discrimination between bulk nuclear recoils and bulk or surface electron recoils while maintaining a large fiducial volume.  

\begin{figure}
   \centering
   \includegraphics[width=0.75\textwidth]{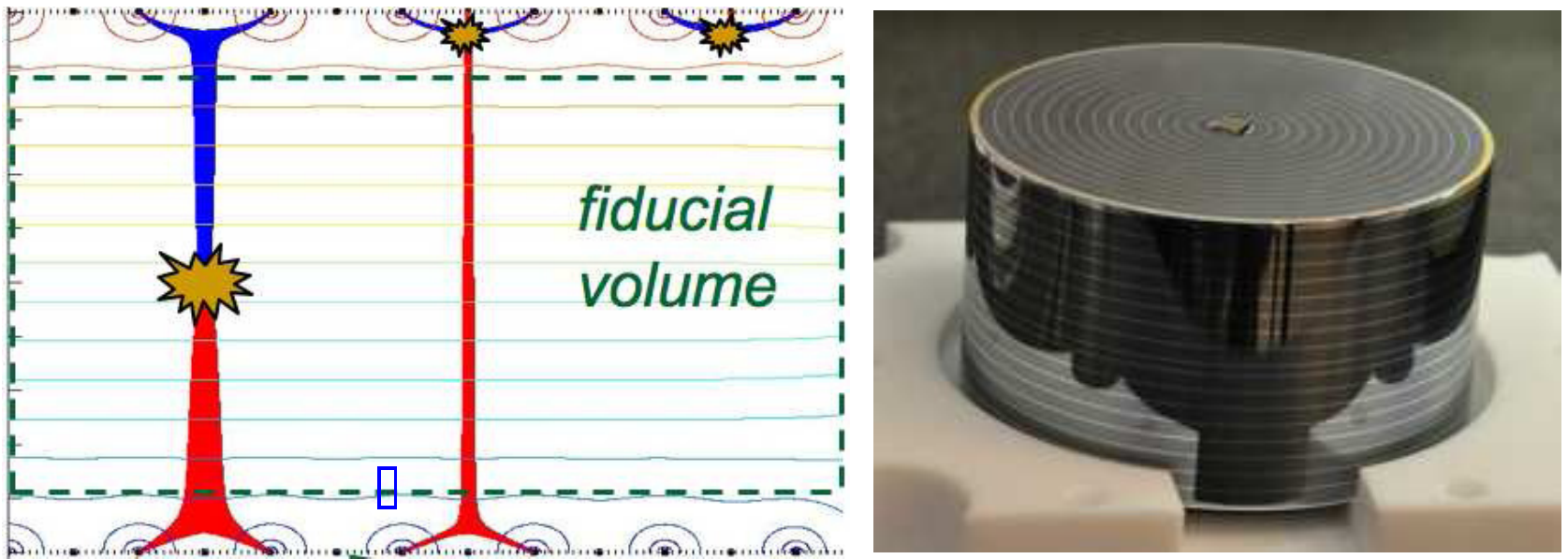}
   \caption{Edelweiss InterDigit detectors. Concentric charge rings biased at three different voltages create a field topology near the surfaces that allows tagging of surface events by looking at the charge partitions. The thermometer used to measure the phonon signal is glued on the middle of the crystal as seen in the photo.}
   \label{fig:edel}
\end{figure}

A similar approach with interleaved electrode sensors, but with a single temperature sensor per detector, is used by the Edelweiss collaboration~\cite{Armengaud:2010wx}. They have placed a very competitive limit based on an effective exposure of 322 kg-day on 9 400~g InterDigit (ID) detectors (see \fref{fig:limits}). A schematic and photo of the latest 800~g version of the InterDigit detectors is shown in \fref{fig:edel}. A set of interleaved electrodes forming concentric rings modifies the electric field topology near the crystal surface, inducing different charge partitions for surface and bulk events. A well-controlled fiducial volume is defined away from the surfaces. The phonon measurement is done by a germanium resistance thermometer attached to the center of the crystal.

The CRESST collaboration operates CaWO$_{4}$ crystal scintillators at cryogenic temperatures, and measures the phonon signal from the crystal along with the scintillation light~\cite{COZZINI:2005it}. The scintillation/phonon ratio allows discrimination between electron and nuclear recoils. CRESST crystals are about 300~g each and are instrumented with a phonon sensor. A separate sapphire wafer with another phonon sensor acts as a photon counter, absorbing the scintillation light which is read out as heat on the sapphire wafer's thermometer. Both crystal and sapphire wafer are enclosed in a common reflective housing to increase the light absorption into the sapphire. 


\section{WIMP--nucleon Limit Plots}\label{sec:Limits}

\begin{figure}
   \centering
   \includegraphics[width=\textwidth]{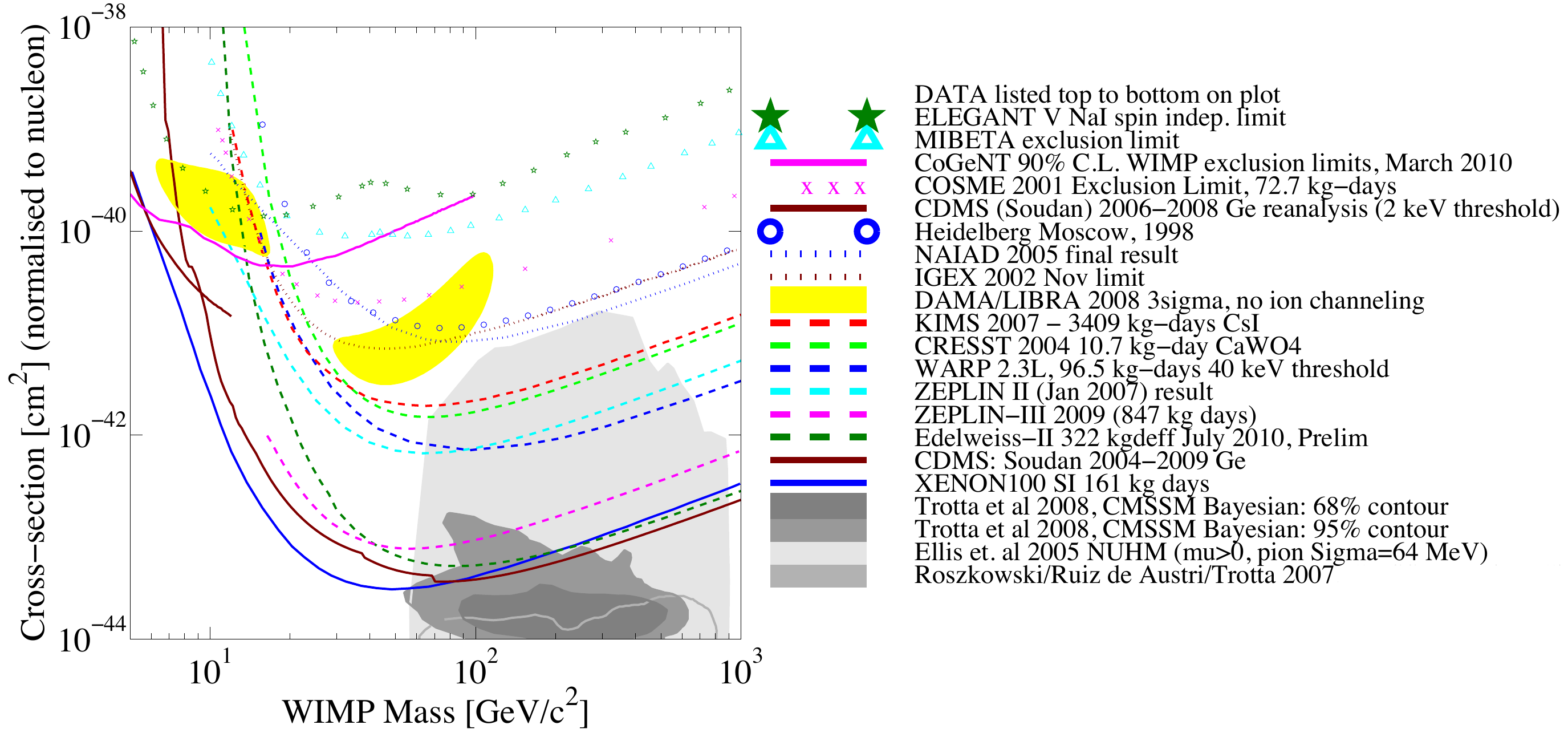}    
   \caption{Current limits on the WIMP--nucleon spin-independent cross section. The region of this model space compatible with the DAMA signal is shown in yellow. Various theoretical predictions are shown in gray tones.}
   \label{fig:limits}
\end{figure}

Having calculated the expected rate, the expected background, and assuming some technique for discrimination has been employed and data has been taken, the final step is to compare the data to the expected WIMP signal. Experiments that do not see a WIMP signal can still place limits on the WIMP--nucleon cross section based on their data by looking at what cross sections \emph{would} have given a detectable signal, and excluding those. The most common method used is the Maximum Interval Method~\cite{2002PhRvD..66c2005Y}. The result is a 90\% confidence limit on the cross section for each potential (and unknown) WIMP mass, as shown in Fig~\ref{fig:limits}. This figure shows most of the published spin-independent WIMP--nucleon cross section limits. The DAMA-compatible regions are shaded yellow, with theoretical predictions in gray.

\bibliographystyle{ws-procs9x6}
\bibliography{tasiref}

\end{document}